\documentclass[apjL]{emulateapj}
\usepackage{graphicx}
\usepackage{rotating}
\usepackage{longtable}
\usepackage{natbib}
\usepackage{latexsym}

\begin{document}

\title{X-ray time variability across the atoll source states of 4U 1636--53}

\author{D. Altamirano\altaffilmark{1}, 
M. van der Klis\altaffilmark{1}, 
M. M\'endez\altaffilmark{1,2} \\
P.G. Jonker\altaffilmark{2,3},  
M. Klein-Wolt\altaffilmark{1} \&    
W.H.G. Lewin\altaffilmark{4} \\
}

\email{d.altamirano@uva.nl}

\altaffiltext{1}{Astronomical Institute, ``Anton Pannekoek'',
University of Amsterdam, and Center for High Energy Astrophysics,
Kruislaan 403, 1098 SJ Amsterdam, The Netherlands.}

\altaffiltext{2}{SRON, Netherlands Institute for Space Research,
Sorbonnelaan 2, 3584 CA, Utrecht, The Netherlands.}

\altaffiltext{3}{Harvard-Smithsonian Center for Astrophysics, 60
Garden Street, MS 83, MA 02138, Cambridge, U.S.A.}

\altaffiltext{4}{MIT Center for Space Research, 70 Vassar Street,
Cambridge, MA 02139}

\date{}

\begin{abstract}
We have studied the rapid X-ray time variability in 149 pointed
observations with the \textit{Rossi X-ray Timing Explorer} (RXTE)'s
Proportional Counter Array of the atoll source 4U~1636--53 in the
banana state and, for the first time with RXTE, in the island state.
We compare the frequencies of the variability components of
4U~1636--53 with those in other atoll and Z-sources and find that
4U~1636--53 follows the universal scheme of correlations
previously found for other atoll sources at (sometimes much) lower
luminosities. 
Our results on the hectohertz QPO suggest  that the mechanism that
sets its frequency differs from that for the other components, while
the amplitude setting mechanism is common.
A previously proposed interpretation of the narrow low-frequency QPO
frequencies in different sources in terms of harmonic mode switching
is not supported by our data, nor by some previous data on other
sources and the frequency range that this QPO covers is found not to
be related to spin, angular momentum or luminosity.

\end{abstract}
\date{} \keywords{accretion, accretion disks --- binaries: close ---
stars: individual (4U 1636--53,4U 1820--30,4U 1608--52,4U 0614+09,4U
1728--34) --- stars: neutron --- X--rays: stars}

\maketitle

\today

\section{Introduction}
\label{sec:intro}

Low-mass X-ray binaries (LMXBs) can be divided into systems containing
a black hole candidate (BHC) and those containing a neutron star
(NS). The accretion process onto these compact objects can be studied
through the timing properties of the associated X-ray emission
\citep[see, e.g., ][for a review]{Vanderklis06}.  \citet{Hasinger89}
classified the NS LMXBs based on the correlated variations of the
X-ray spectral and rapid X-ray variability properties. They
distinguished two sub-types of NS LMXBs, the Z sources and the atoll
sources, whose names were inspired by the shapes of the tracks that
they trace out in an X-ray color-color diagram on time scales of hours
to days. The Z sources are the most luminous;  the atoll sources
cover a much wider range in luminosities \citep[e.g. ,][and references
therein]{Ford00}. For each type of source, several spectral/timing
states are identified which are thought to arise from qualitatively
different inner flow configurations.  In the case of atoll sources,
the main three states are the extreme island state (EIS), the island
state (IS) and the banana branch, the latter subdivided into
lower-left banana (LLB), lower banana (LB) and upper banana (UB)
states.  Each state is characterized by a unique combination of color
color diagram and timing behavior.
The EIS and the IS occupy the spectrally harder parts of the color
color diagram (CD) corresponding to lower X-ray luminosity ($L_x$).
The different patterns they show in the CD are traced out in days to
weeks.  The hardest and lowest $L_x$ state is generally the
\textit{EIS}, which shows strong low-frequency flat-topped noise.  The
\textit{IS} is spectrally softer than the \textit{EIS}.  Its
power spectrum is characterized by broad features and a dominant
band-limited noise (BLN) component which becomes stronger and lower in
characteristic frequency as the flux decreases and the $>6$ keV
spectrum gets harder.  In order of increasing $L_x$ we encounter the
\textit{LLB}, where the twin kHz QPOs are first observed, the
\textit{LB}, where dominant 10-Hz BLN occurs and finally, the
\textit{UB}, where the $<1$~Hz (power law) very low frequency noise
(VLFN) dominates.  In the banana states, some of the broad features
observed in the EIS and the IS become narrower (peaked) and occur at
higher frequency. The twin kHz QPOs can be found in
\textit{LLB} at frequencies in excess of  1000~Hz, only one is seen in the
\textit{LB}, and no kHz QPOs are detected in the \textit{UB} \citep[see
reviews by][for detailed descriptions of the different
states]{Hasinger89,Vanderklis00,Vanderklis04,Vanderklis06}.

4U~1636--53 is an atoll source \citep{Hasinger89} which has an orbital
period of $\sim3.8$ hours \citep{Vanparadijs90} and a companion star
with a mass of $\sim0.4 \ M_{\odot}$ \citep[assuming a NS of $\sim1.4
\ M_{\odot}$, see][for a discussion]{Giles02}.  It was first observed
as a strong continuous X-ray source (Norma X-1) with Copernicus
\citep{Willmore74} and Uhuru \citep{Giacconi74}.  4U~1636--53 is an
X-ray burst source \citep{Hoffman77} which shows asymptotic burst
oscillation frequencies of $\sim581$~Hz \citep[see
e.g.][]{Zhang97,Giles02}.  This is probably the approximate spin
frequency; although \citet{Miller99} presented evidence that these
oscillations might actually be the second harmonic of a neutron star
spin frequency of $\sim290$~Hz, this was not confirmed in further work
by \citet{Strohmayer01}.  \citet{Prins97} studied the aperiodic timing
behavior of 4U~1636--53 with the EXOSAT Medium Energy instrument up to
frequencies of $\sim100$~Hz both in the island and the banana
state. \citet{Wijnands97}, using observations with RXTE, discovered
two simultaneous quasi-periodic oscillations (QPOs) near 900~Hz and
1176~Hz when the source was in the banana state.  The frequency
difference $\Delta\nu$ between the two kHz QPO peaks is nearly equal
to half the burst oscillation frequency, similar to what has been
observed in other sources with burst oscillations or pulsation
frequency $>400$~Hz.  To the extent that this implies $\Delta\nu \sim
\nu_{spin}/2$, this is inconsistent with spin-orbit beat-frequency
models \citep{Wijnands03} for the kHz QPOs such as proposed by
\citet{Miller98}.  Other complications for beat frequency models
include the fact that $\Delta\nu$ is neither constant \citep[e.g. in
Sco X-1, ][]{Vanderklis97} nor exactly equal to half the burst
oscillation frequency.  Generally, $\Delta\nu$ decreases as the kHz
QPO frequency increases, and in 4U~1636--53, observations have shown
$\Delta\nu$ at frequencies lower as well as higher than half the burst
oscillation frequency \citep{Mendez98,Jonker02}.

\citet{Straaten02,Straaten03} compared the timing properties of
4U~0614+09, 4U~1608--52 and 4U~1728--34 and conclude that the
frequencies of the variability components in these sources follow the
same pattern of correlations.  \citet{Disalvo03}, based on five
detections of kHz QPOs in 4U~1636--53 in the banana state was able to
show that at least in that state the source might fit in with that
same scheme of correlations. 
The detailed investigation of 4U~1636--53 is important because it is
one of the most luminous atoll sources \citep{Ford00} that shows the
full complement of island (this paper) and banana states 
and that also shares other timing features with often less luminous atoll
sources. For example, 
\citet{Revnivtsev01} found a new class of low frequency QPOs in the
mHz range which they suggested to be associated with nuclear burning
in 4U~1636--53 and 4U~1608--52.  \citet{Mendez00a} and
\citet{Mendez01} compared the relations between kHz QPOs and inferred
mass accretion rate in 4U~1728--34, 4U~1608--52, Aql~X-1 and
4U~1636--53, and showed that the dependence of the frequency of one of
the kHz QPOs upon X-ray intensity is complex, but similar among
sources.  \citet{Jonker00} discovered a third kHz QPO in 4U~1608--52,
4U~1728--34, and 4U~1636--53 which is likely an upper sideband to the
lower kHz QPO.  Recently, \citet{Jonker05} found in 4U~1636--53 an
additional (fourth) kHz QPO, likely the corresponding lower sideband.

In this paper, we present new results for low frequency noise with
characteristic frequencies $1-100$~Hz and QPOs in the range $100-1260$
Hz, for the first time including RXTE observations of the island state
of this source.  
These results better constrain the timing behavior in the various
states of 4U~1636--53.  We compare our results mainly with those of
the atoll sources 4U~0614+09, 4U~1608--52 and 4U~1728--34 and find
that the frequency of the hectohertz component may not be constant as
previously stated, but may have a sinusoidal like modulation within
its range from $\sim100$ to $\sim250$~Hz.  Our results also
suggest that the mechanism that sets the frequency of the hHz QPOs
differs from that for the other components, while the amplitude
setting mechanism is common.
Finally, we demonstrate that it is not
possible to clearly distinguish between two harmonics of the
low-frequency QPO $L_{LF}$ across different sources, as was previously
thought \citep{Straaten03}.

\begin{figure*}[!hbtp] 
\center
\resizebox{0.7\columnwidth}{!}{\rotatebox{0}{\includegraphics{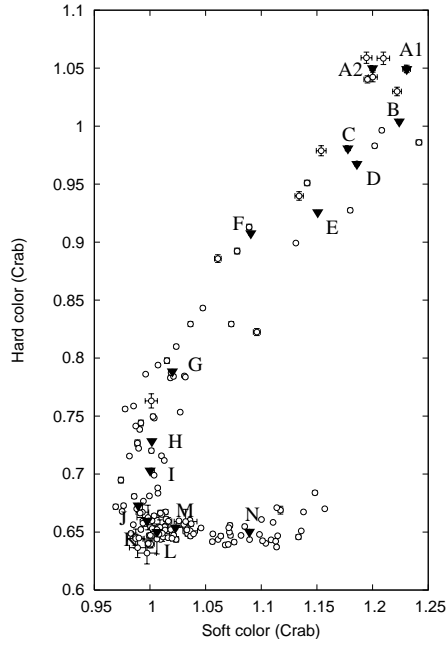}}}
\caption{Hard color versus soft color normalized to the Crab Nebula as explained in Section~\ref{sec:intro}. 
Each circle represents the average soft/hard color of one of the observations used for this paper. 
The filled triangles mark averages of 1 to 32 observations and are labeled with letters, in order
from ($A$) the island state, Lower Left Banana ($J$) to the Lower banana ($N$). 
}
\label{fig:ccd}
\end{figure*}

\begin{figure*}[!hbtp] 
\center
\resizebox{0.7\columnwidth}{!}{\rotatebox{0}{\includegraphics{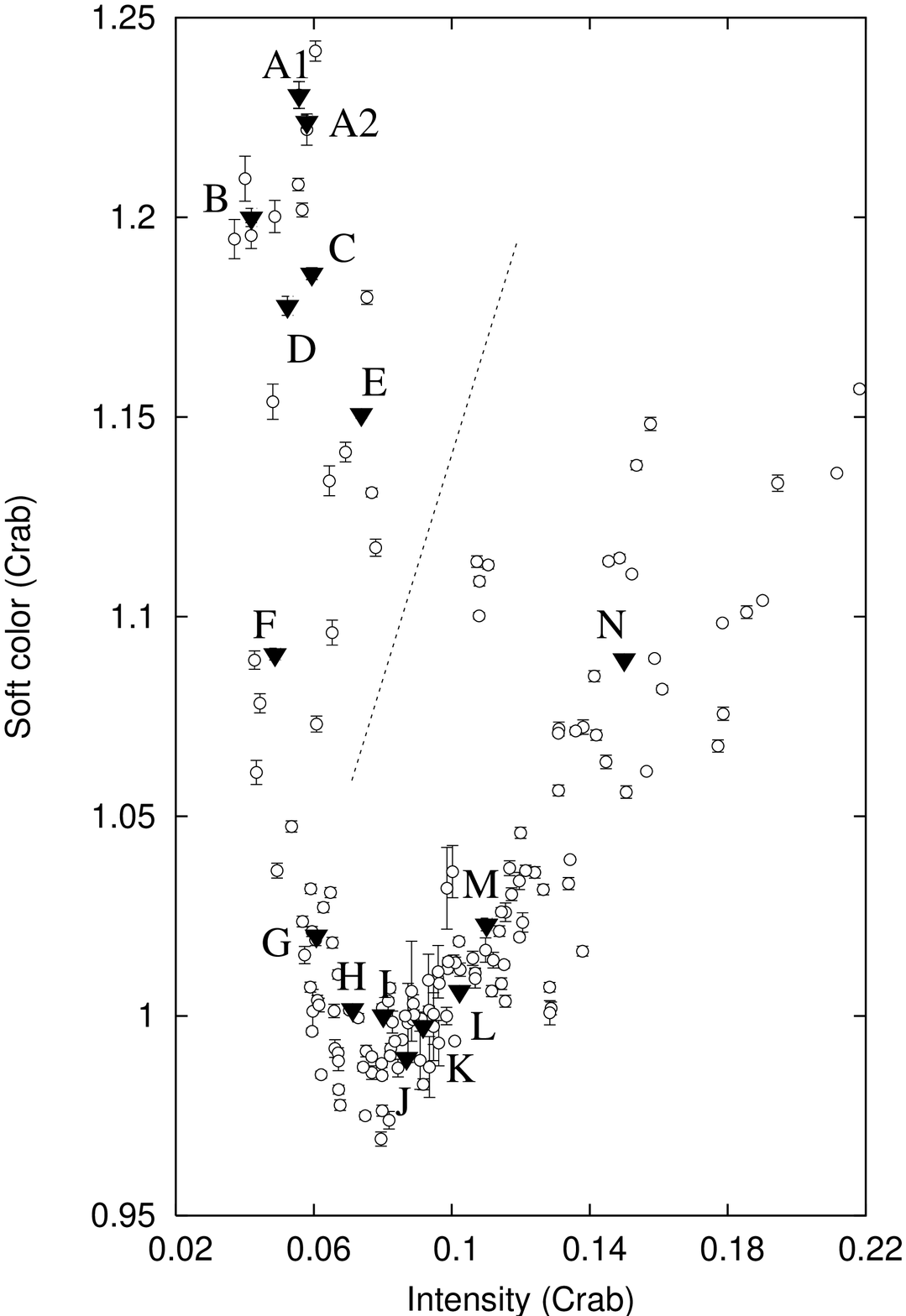}}}
\resizebox{0.7\columnwidth}{!}{\rotatebox{0}{\includegraphics{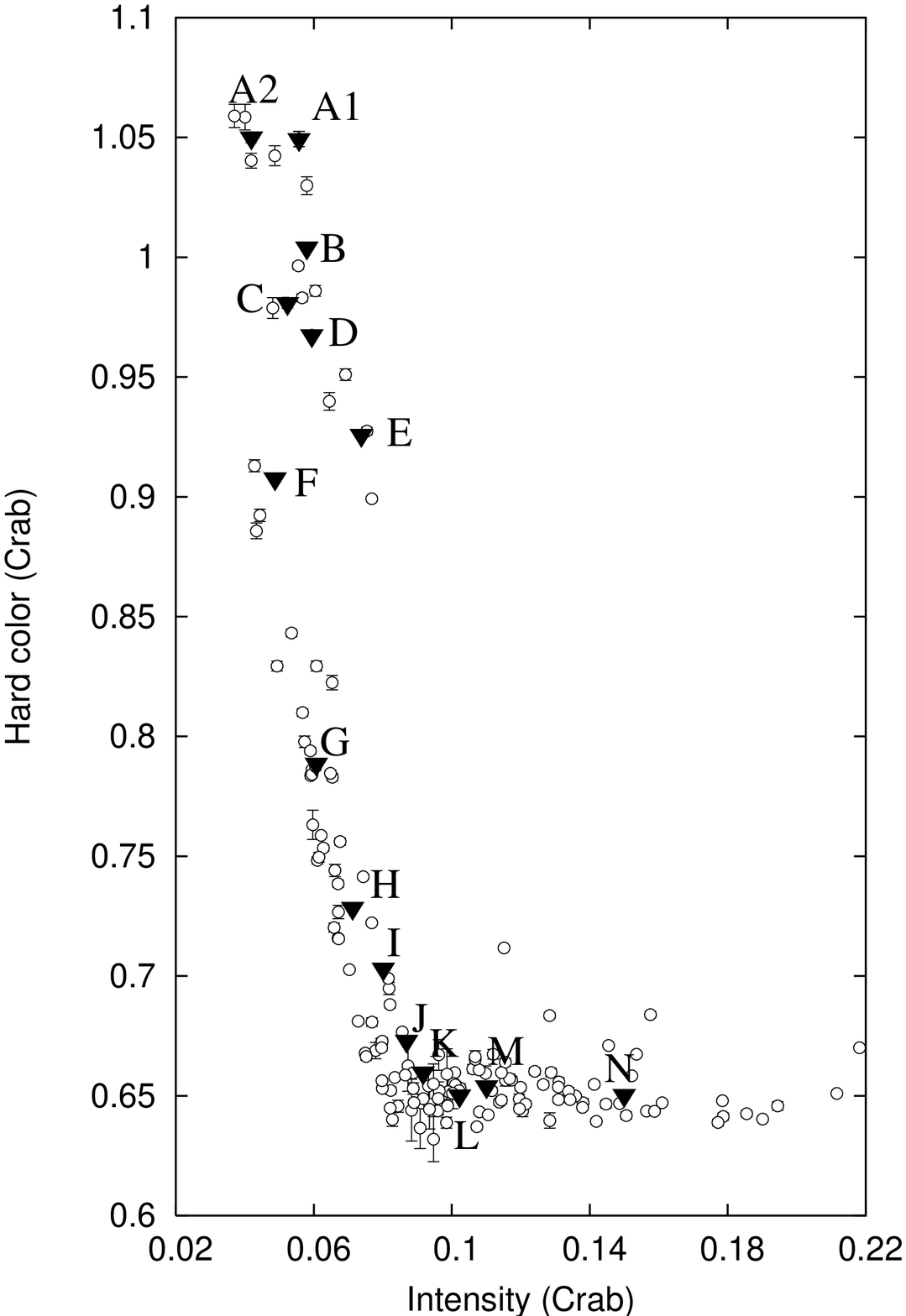}}}
\caption{Soft color vs. intensity (left) and hard color vs. intensity
(right) in units of the Crab Nebula as explained in
Section~\ref{sec:intro}. Symbols as in Figure~\ref{fig:ccd}. 
For clarity, the dashed line 
separates the observations corresponding to the IS (left), from the observations
corresponding to the BS (right).
}
\label{fig:cvsint}
\end{figure*}

\section{OBSERVATIONS AND DATA ANALYSIS}
\label{sec:dataanalysis}

We use data from the Rossi X-ray Timing Explorer (RXTE) Proportional
Counter Array \citep[PCA; for instrument information
see][]{Zhang93}. There were 149 pointed observations in the four data
sets we used (60032-01, 60032-05, 70036-01, 80425-01 \& 90409-01),
each consisting of a fraction of one to several entire satellite
orbits, for $\sim1$ to $\sim26$ ksec of useful data per observation.
We use the 16-s time-resolution Standard 2 mode data to calculate
X-ray colors as described in \citet{Altamirano05}.  Hard and soft
color are defined as the 9.7--16.0~keV~/~6.0--9.7~keV and
3.5--6.0~keV~/~2.0--3.5~keV count rate ratio, respectively, and
intensity as the 2.0--16.0~keV count rate.  Type I X-ray bursts were
removed, background was subtracted and deadtime corrections were made.
In order to correct for the gain changes as well as the differences in
effective area between the PCUs themselves, we normalize our colors by
the corresponding Crab color values
\cite[see][]{Kuulkers94,Straaten03} that are closest in time but in
the same RXTE gain epoch, i.e., with the same high voltage setting of
the PCUs \citep{Jahoda05}. In Table~\ref{table:crab} we show for
reference the all-epoch averaged colors for Crab Nebula. 
 All active PCUs were used to calculate the
colors in 4U~1636--53 except for observation 60032-01-01-02, where due
to a PCU3 malfunction we only used PCUs 0 and 2.
Figure~\ref{fig:ccd} shows the color-color diagram of the 149
different observations that we used for this analysis, and
Figure~\ref{fig:cvsint} the corresponding hardness-intensity diagrams
(soft and hard color vs. intensity).

For the Fourier timing analysis we used data from the $\sim125\mu$s
(1/8192 s) time resolution Event mode E\_125us\_64M\_0\_1s.  First, we
used a 2 second-binned light curve in order to detect and remove data
drop-outs and X-ray bursts (these data were also excluded from
the rest of the analysis).  Leahy-normalized power spectra were
constructed using data segments of 128 seconds and 1/8192s time bins
such that the lowest available frequency is $1/128 \approx 8 \times
10^{-3}$~Hz and the Nyquist frequency 4096~Hz.  No background or
deadtime corrections were made prior to the calculation of the
power spectra.  We first averaged the power spectra per observation.
We inspected the shape of the average power spectra at high frequency
( $>2000$~Hz) for unusual features in addition to the usual Poisson
noise. None were found.  We then subtracted a Poisson noise spectrum
estimated from the power between 3000 and 4000~Hz, where neither
intrinsic noise nor QPOs are known to be present, using the method
developed by \citet{Kleinwolt04} based on the analytical function of
\citet{Zhang95}.  The resulting power spectra were converted to
squared fractional rms \citep{Vanderklis95b}.  In this normalization
the square root of the integrated power density equals the variance of
the intrinsic variability in the source count rate. 
In order to improve the statistics, observations were averaged
 together if they described the same source state. Since it is known
 from previous work on similar sources that the position of the source
 in the color-color diagram generally is well correlated to its
 spectral/timing state \citep[see e.g.][ and references
 within]{Vanderklis06}, we first grouped observations with similar
 colors.  Within each group, we then compared the shape of each
 average power spectrum with all of the other ones to create subgroups
 in which all power spectra had a dependence of power on frequency
 that was identical within errors. So, narrow features had to be at
 the same frequency for average power spectra to be added together.
 The resulting data selections are labeled interval A to N (see
 Table~\ref{table:averages} for details on which observations were
 used for each interval and their colors).
 A disadvantage of this method is that we can loose information about
 narrow features moving on time scales shorter than an observation,
 such as the lower kilohertz QPO \citep[see
 e.g.][]{Berger96,Disalvo03}.
The ``shift and add'' method \citep{Mendez98a}, to some extent might
 be able to compensate for this; we explore in the Appendix this
 issue.
Our method is the best suited one to study the behavior of the broad features
such as typically seen in low mass X-ray binaries' power spectra
\citep[e.g.][]{Straaten02,Straaten03,Straaten05,Altamirano05,Manu05}.
For these broad components, which are the main aim of this paper, the
 gain in signal to noise due to this averaging process outweighs a
 minor additional broadening due to frequency variations.

\begin{table*}[!hbtp]
\center
\begin{tabular}{|c|c|c|c|}\hline 
\multicolumn{4}{|c|}{Interval A1} \\
\hline
 Observation  & Soft color (Crab) & Hard color (Crab) & Intensity (Crab) \\
\hline
80425-01-04-01 & $1.2306\pm0.0034$ & $1.0493\pm0.0032$ & $0.0557\pm0.0001$ \\
\hline
\multicolumn{4}{|c|}{Interval A2} \\
\hline
80425-01-03-00 & $1.2097\pm0.0056$ & $1.0585\pm0.0053$ & $0.0400\pm0.0001$ \\ 
90409-01-01-00 & $1.1945\pm0.0049$ & $1.0590\pm0.0049$ & $0.0370\pm0.0001$ \\ 
90409-01-01-01 & $1.1954\pm0.0032$ & $1.0403\pm0.0031$ & $0.0418\pm0.0001$ \\ 
90409-01-02-00 & $1.2002\pm0.0040$ & $1.0424\pm0.0042$ & $0.0487\pm0.0001$ \\ 
\hline
\end{tabular}
\caption{Observations used for the timing analysis. The colors and intensity are normalized
to Crab (See Section~\ref{sec:dataanalysis}).
The complete table can be obtained digitally from ApJ.}
\label{table:averages}
\end{table*}

\begin{table*}[!hbtp]
\center
\begin{tabular}{|c|c|c|c|}\hline 
\hline
 PCU Number & Crab's soft color  & Crab's hard color & Crab's intensity (c/s) \\
\hline
0 & $2.21\pm0.04$ & $0.56\pm0.009$ & $2552\pm21$ \\
1 & $2.27\pm0.03$ & $0.55\pm0.008$ & $2438\pm21$ \\
2 & $2.22\pm0.04$ & $0.58\pm0.010$ & $2424\pm20$ \\
3 & $2.42\pm0.03$ & $0.57\pm0.009$ & $2365\pm20$ \\
4 & $2.34\pm0.03$ & $0.58\pm0.011$ & $2299\pm21$ \\
\hline
\end{tabular}
\caption{Average soft, hard and intensity of the Crab Nebula over all
epochs and per PCU. Note that we have used averaged values \textit{per
day} in our analysis; the numbers listed here are representative for
those daily averages.  We quoted the averaged quadratic errors,
i.e. $\sqrt{\sum_i^n\Delta\kappa^2_i}/n$, where $\Delta\kappa$ is
either the soft, hard or the intensity day error. }
\label{table:crab}
\end{table*}

\begin{figure}[!hbtp]
\centering 
\resizebox{0.50\columnwidth}{!}{\rotatebox{0}{\includegraphics{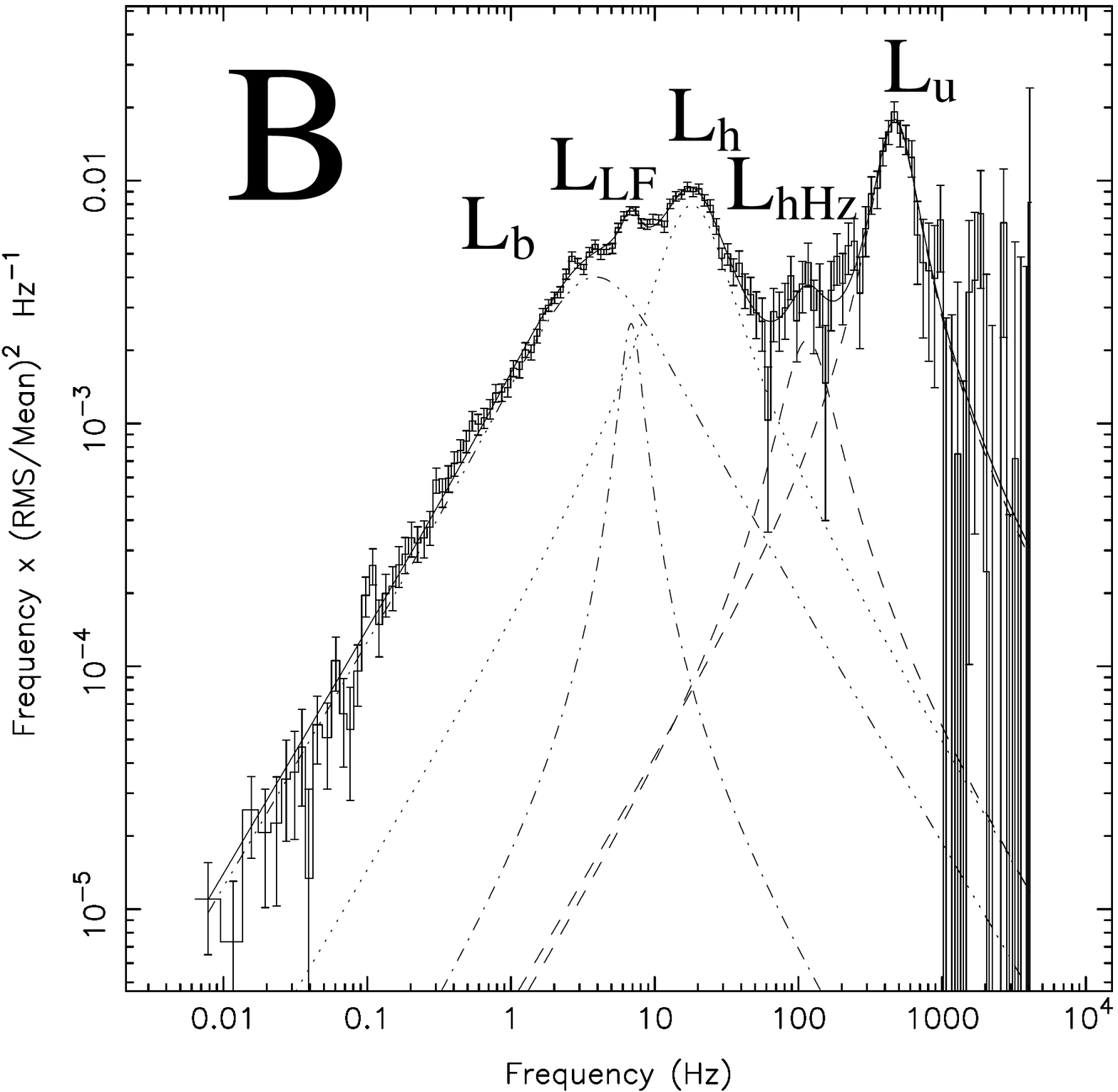}}} 
\resizebox{0.50\columnwidth}{!}{\rotatebox{0}{\includegraphics{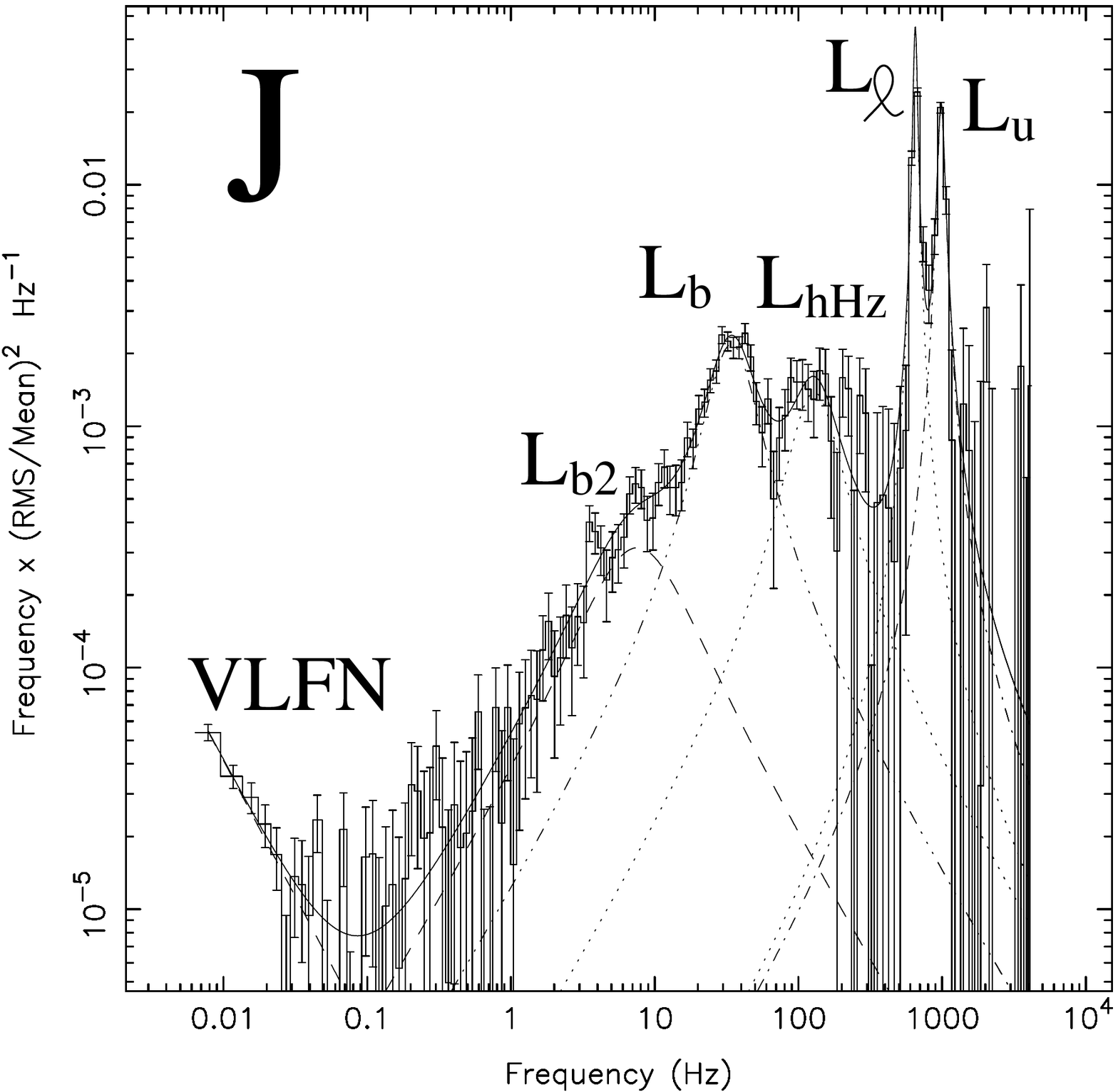}}}          
\caption{Representative power spectra of the island state (above - interval B) and
the banana state (below - interval J) with their components.}
\label{fig:components}
\end{figure}

To fit the power spectra, we used a multi-Lorentzian function: the sum
of several Lorentzian components plus, if necessary, a power law to
fit the very low frequency noise at $\lesssim1$~Hz. Each Lorentzian
component is denoted as $L_i$, where $i$ determines the type of
component. The characteristic frequency ($\nu_{max}$ as defined below)
of $L_i$ is denoted $\nu_i$.  For example, $L_u$ identifies the upper
kHz QPO and $\nu_u$ its characteristic frequency.  By analogy, other
components have names such as $L_{\ell}$ (lower kHz), $L_{hHz}$
(hectohertz), $L_h$ (hump), $L_b$ (break frequency), and their
frequencies are $\nu_{\ell}$, $\nu_{hHz}$, $\nu_h$ and $\nu_b$,
respectively. For reference, in Figure~\ref{fig:components} we show
two representative power spectra in which we labeled the different
components.  Using this multi-Lorentzian function makes it
straightforward to directly compare the different components in
4U~1636--53 to those in previous works which used the same fit
function \citep[e.g.,][and references
therein]{Belloni02,Straaten02,Straaten03,Straaten05,Altamirano05}.

We only include those Lorentzians in the fits whose single trial
significance exceeds $3\sigma$ based on the negative error bar in the
power integrated from 0 to $\infty$ 
(i.e. we include only those Lorentzians whose integral power
is at least 3 times higher than zero based on the negative 1$\sigma$
error) and  whose inclusion gives a $>3\sigma$ improvement of the fit
according to an F-test.
We give the frequency of the Lorentzians in terms of characteristic
frequency $\nu_{max}$ as introduced by \citet{Belloni02}: $\nu_{max} =
\sqrt{\nu^2_0 +(FWHM/2)^2} = \nu_0 \sqrt{1 + 1/4Q^2}$.  For the
quality factor $Q$ we use the standard definition $Q =
\nu_0/FWHM$. FWHM is the full width at half maximum and $\nu_0$ the
centroid frequency of the Lorentzian. Note that Q values in excess of
$\sim3$ will generally be affected by smearing in an analysis such as
ours. Such values are commonly seen in $L_{LF}$, $L_{\ell}$ and
$L_u$. In Section~\ref{sec:results} we indicate in which cases this
could have occurred.

We only report the results for $\nu_{max} \gtrsim 1$ Hz. 4U~1636--53
is one of three atoll sources which are known to show milihertz QPOs
which affect the power law behavior of the noise at $\lesssim1$ Hz
\citep{Revnivtsev01}.  A different kind of analysis is needed to study
these QPOs; we will report the results in a separate paper
\citep{Altamirano08}.

\begin{table*}[!hbtp]
\center
\begin{tabular}{|ccccccc|}\hline 
\hline
Interval& $\nu_{u0}$       & $FWHM_{\nu_{u0}}^*$ & $\nu_{\ell0}$  & $FWHM_{\nu_{\ell0}}^*$ & $\Delta\nu_0$ & Ratio $\nu_{u0}/\nu_{\ell0}$\\
\hline
H      & $860.4\pm1.7$ & $136.2\pm4.8$ & $565.4\pm5.1$ & $90.4\pm12.3$     & $295.0\pm5.4$ & $1.52\pm0.01$ \\
I      & $896.8\pm2.6$ & $115.8\pm6.3$ & $580.4\pm5.1$ & $77.2\pm12.5$     & $316.4\pm5.7$ & $1.54\pm0.01$ \\
J      & $972.1\pm3.5$ & $98.0\pm8.7$  & $661.1\pm1.8$ & $73.8\pm4.1$      & $311.0\pm3.9$ & $1.470\pm0.006$ \\
K      & $992.9\pm6.6$ & $177.4\pm16.9$ & $718.3\pm2.4$ & $121.5\pm6.3$     & $274.6\pm5.7$ & $1.38\pm0.01$ \\
L      & $1147.2\pm16.6$ & $123.4\pm27.8$ & $836.1\pm0.7$ & $64.1\pm1.9$     & $311.1\pm16.6$ & $1.37\pm0.02$\\
\hline
\hline
\end{tabular}
\caption{Central frequencies, full width at half maximum (FWHM) and
 the frequency difference $\Delta\nu_0$ between the kHz QPOs for the 5
 intervals where both kHz QPOs where detected significantly
 ($>3\sigma$).  $^*$: Note that these values might have been affected
 by smearing (see text) .}
\label{table:centralfreq}
\end{table*}

\section{Results}\label{sec:results}

Figures \ref{fig:ccd} and \ref{fig:cvsint} show that in order A to H,
the spectrum becomes softer (i.e. hard and soft color both decrease),
and the intensity changes little.  From H to L the soft color remains
approximately constant but above 6 keV the spectrum becomes even
steeper (i.e., the hard color decreases further) and, from interval G,
the intensity increases.  Finally from L to N, below 6 keV the
spectrum becomes flatter and above 6 keV it remains approximately
constant in slope, while the intensity continues increasing.  Similar
behavior has been observed in other atoll sources which are moving
from the island to the lower left banana and then to the lower banana
state \citep[see for example][]{Straaten03,Altamirano05}.

\begin{figure*}[!hbtp]
\centering 
\resizebox{0.4\columnwidth}{!}{\rotatebox{0}{\includegraphics{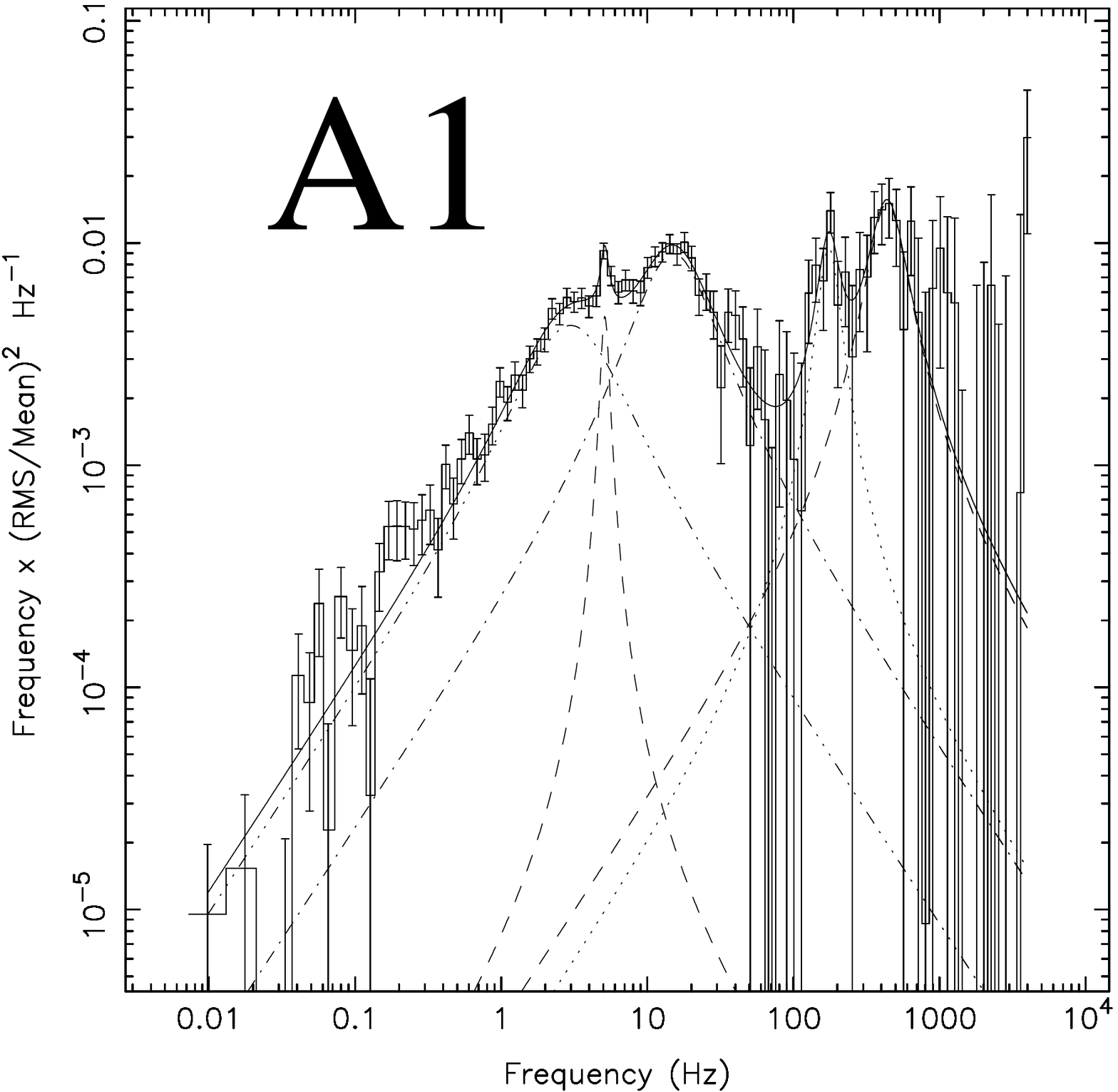}}}                
\resizebox{0.4\columnwidth}{!}{\rotatebox{0}{\includegraphics{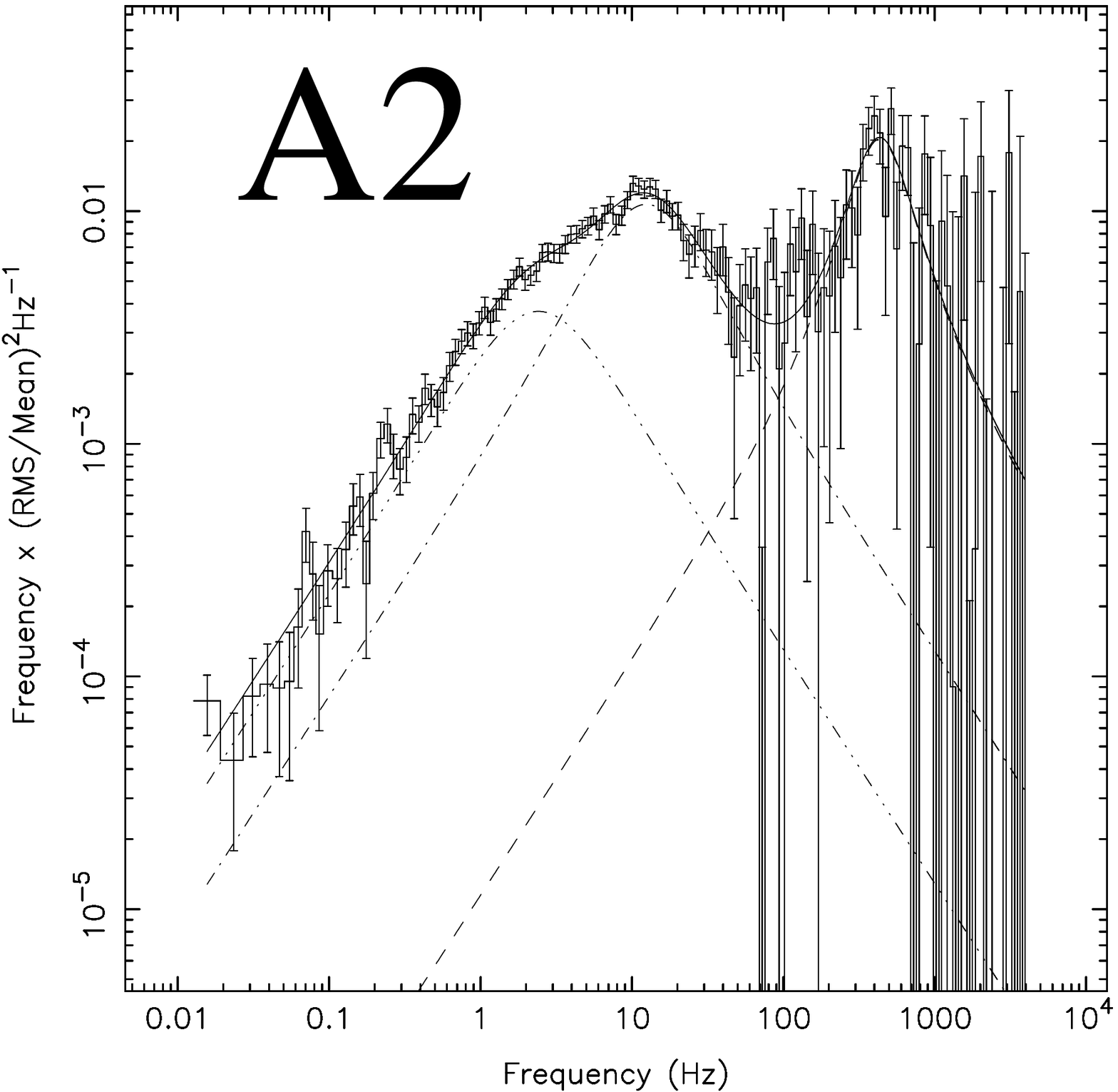}}}                
\resizebox{0.4\columnwidth}{!}{\rotatebox{0}{\includegraphics{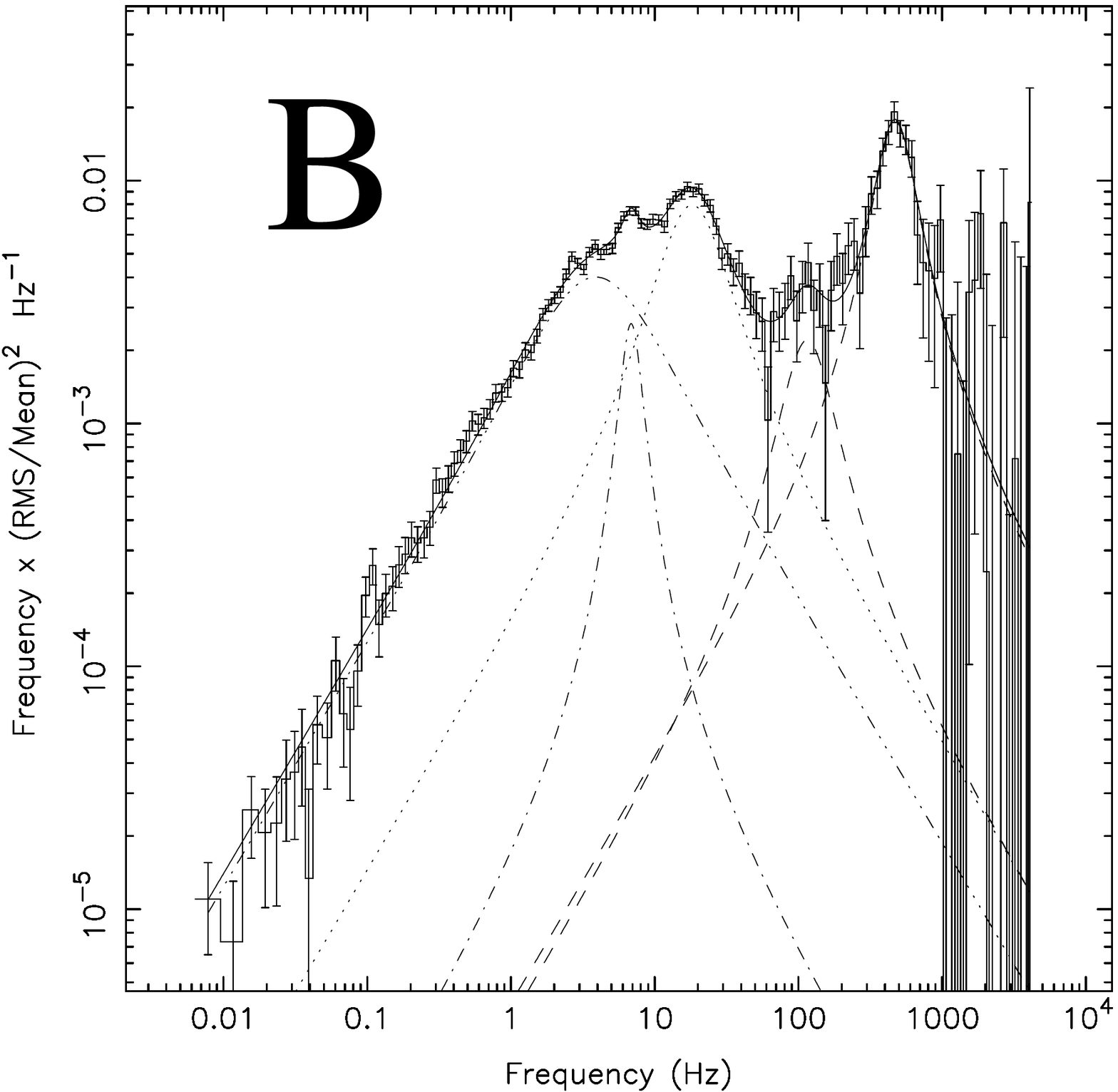}}}                
\resizebox{0.4\columnwidth}{!}{\rotatebox{0}{\includegraphics{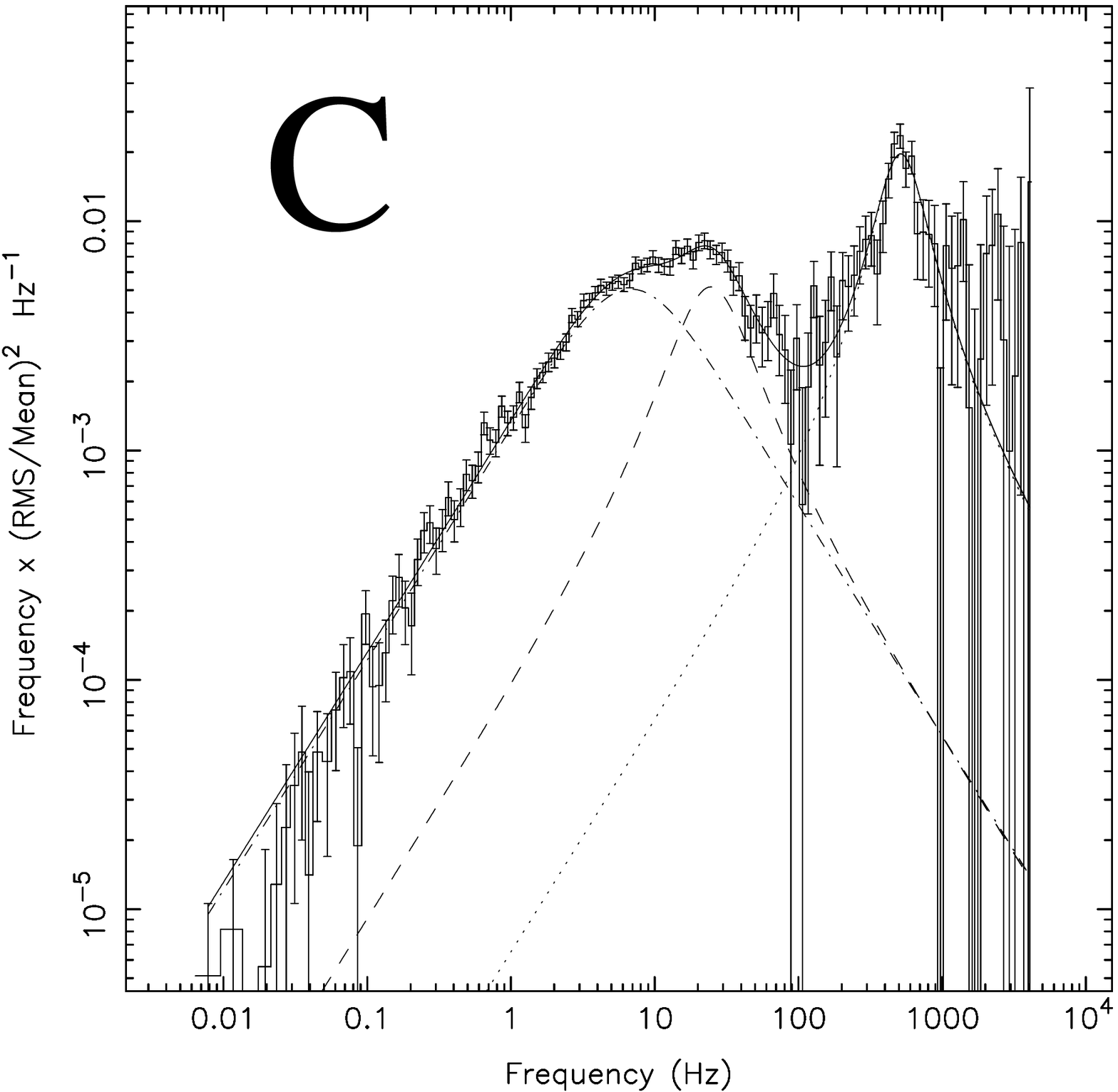}}}                   
\resizebox{0.4\columnwidth}{!}{\rotatebox{0}{\includegraphics{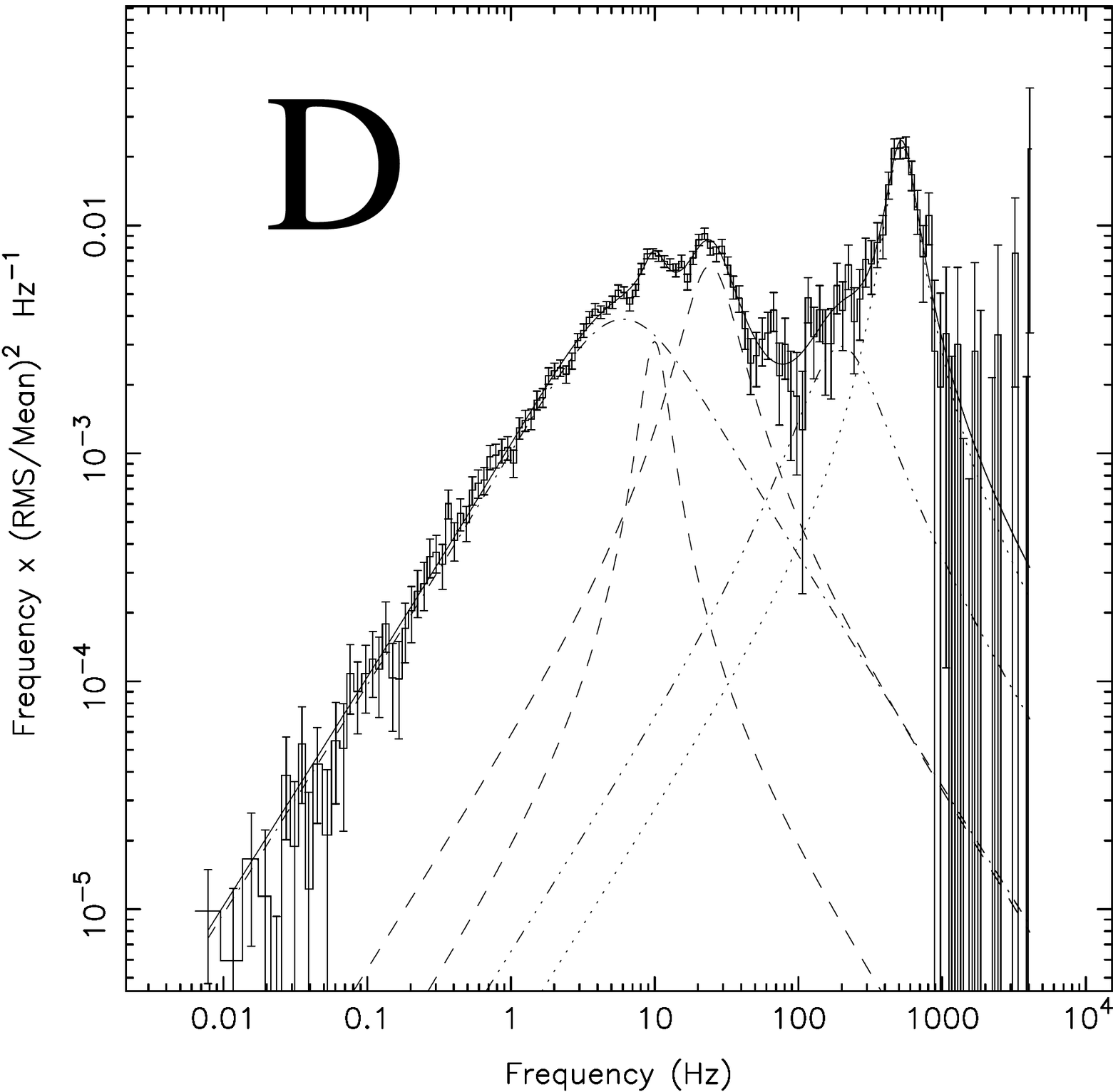}}}                   
\resizebox{0.4\columnwidth}{!}{\rotatebox{0}{\includegraphics{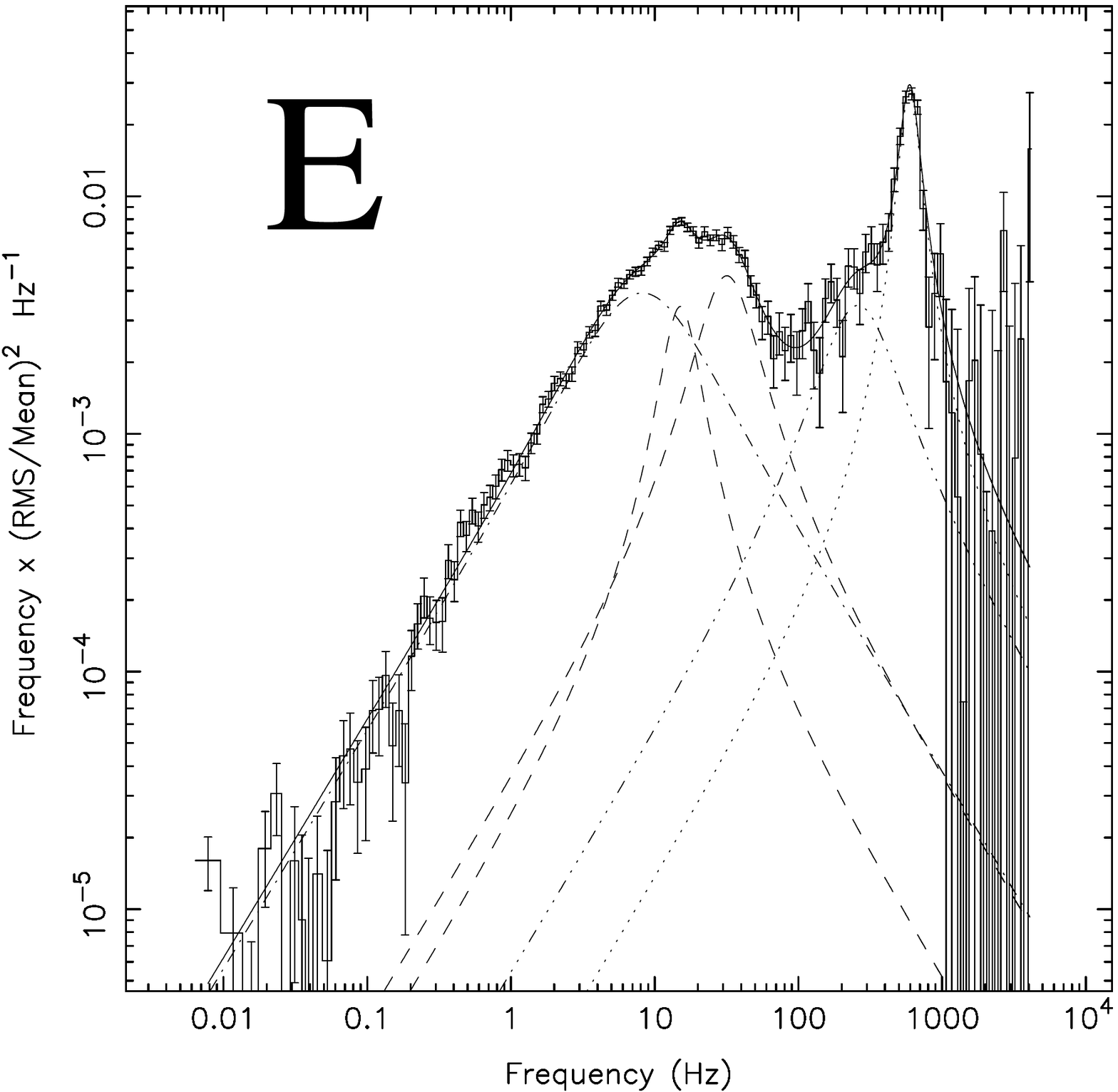}}}                
\resizebox{0.4\columnwidth}{!}{\rotatebox{0}{\includegraphics{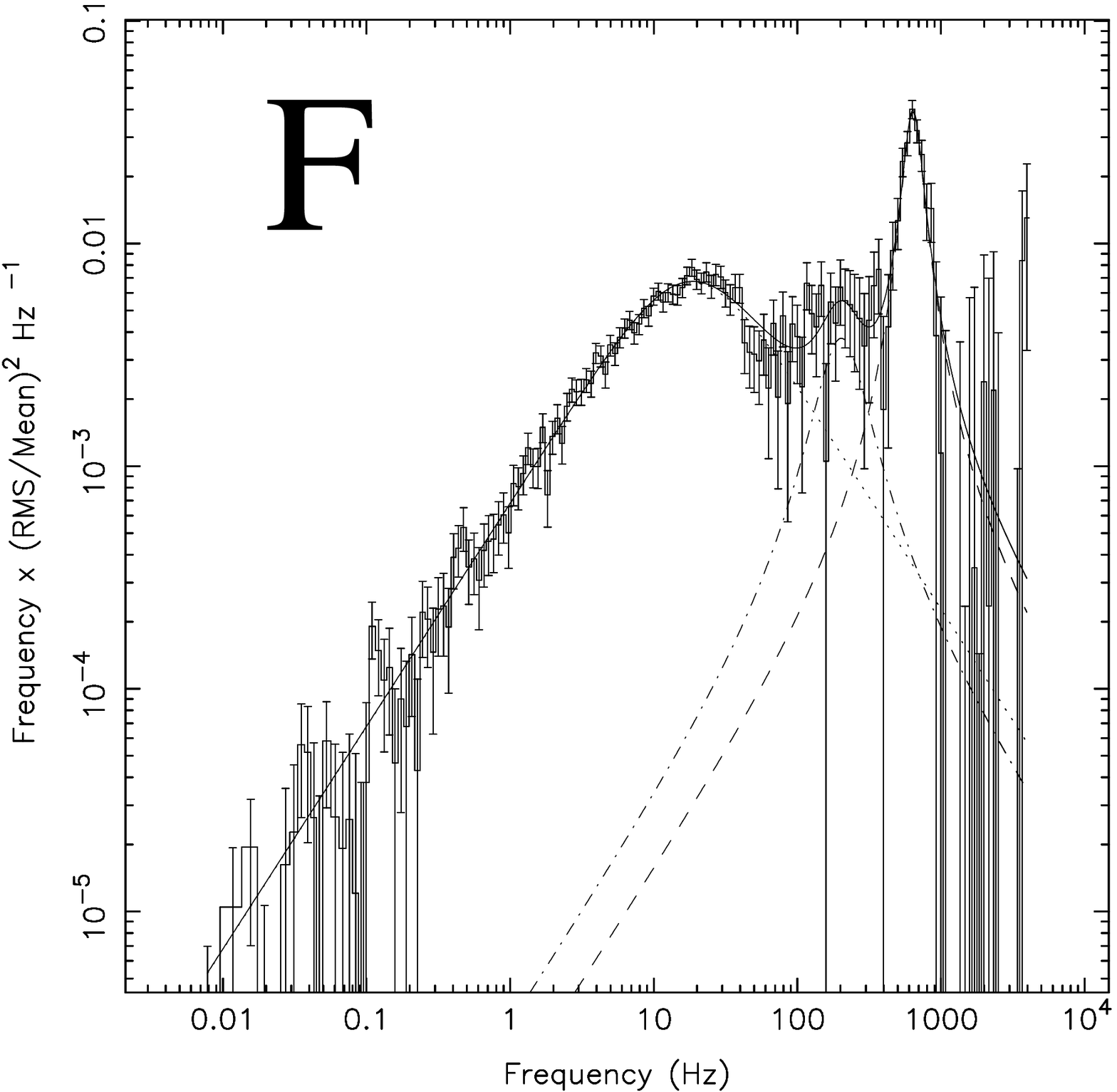}}}                          
\resizebox{0.4\columnwidth}{!}{\rotatebox{0}{\includegraphics{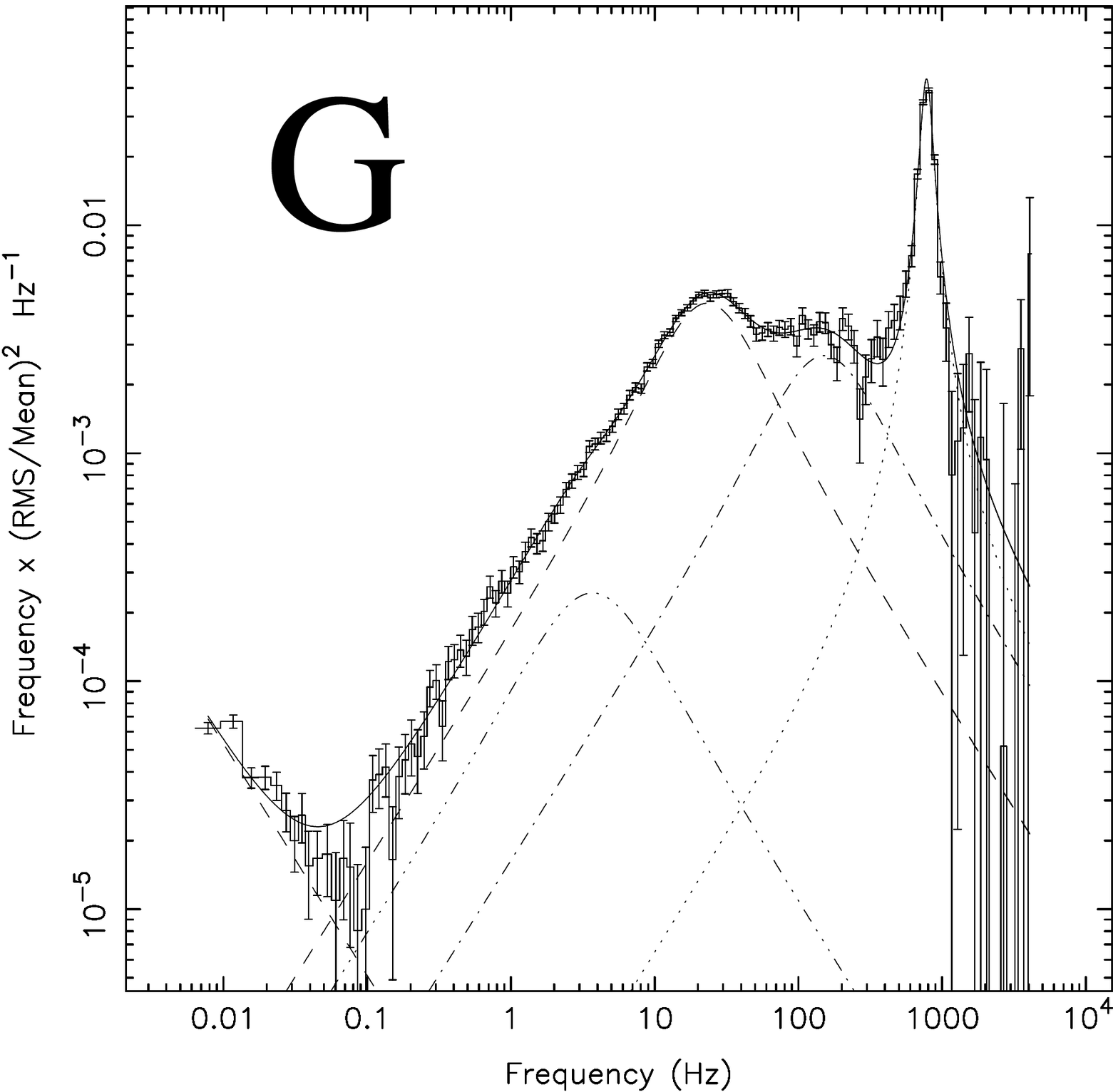}}}                     
\resizebox{0.4\columnwidth}{!}{\rotatebox{0}{\includegraphics{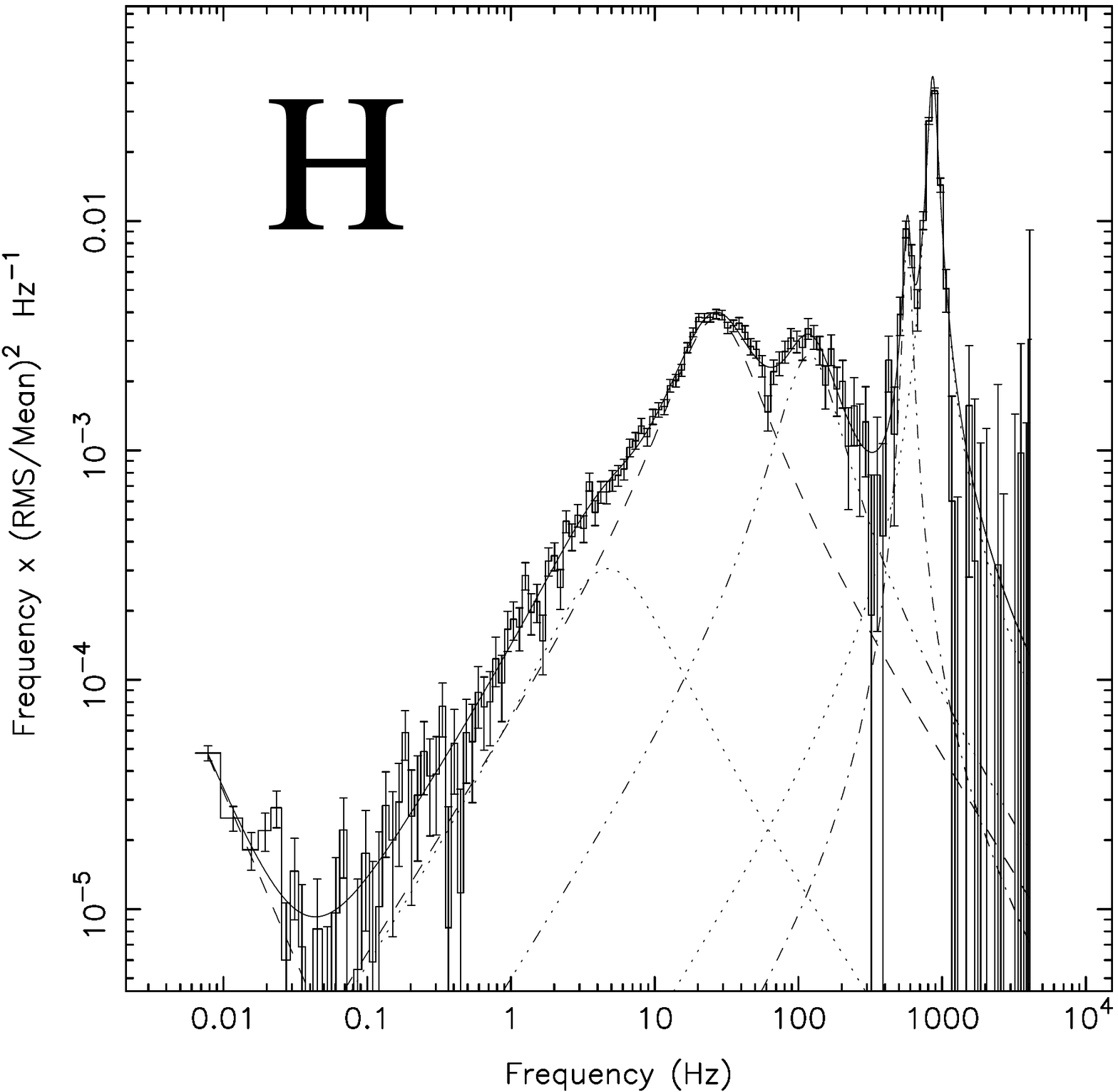}}}                    
\resizebox{0.4\columnwidth}{!}{\rotatebox{0}{\includegraphics{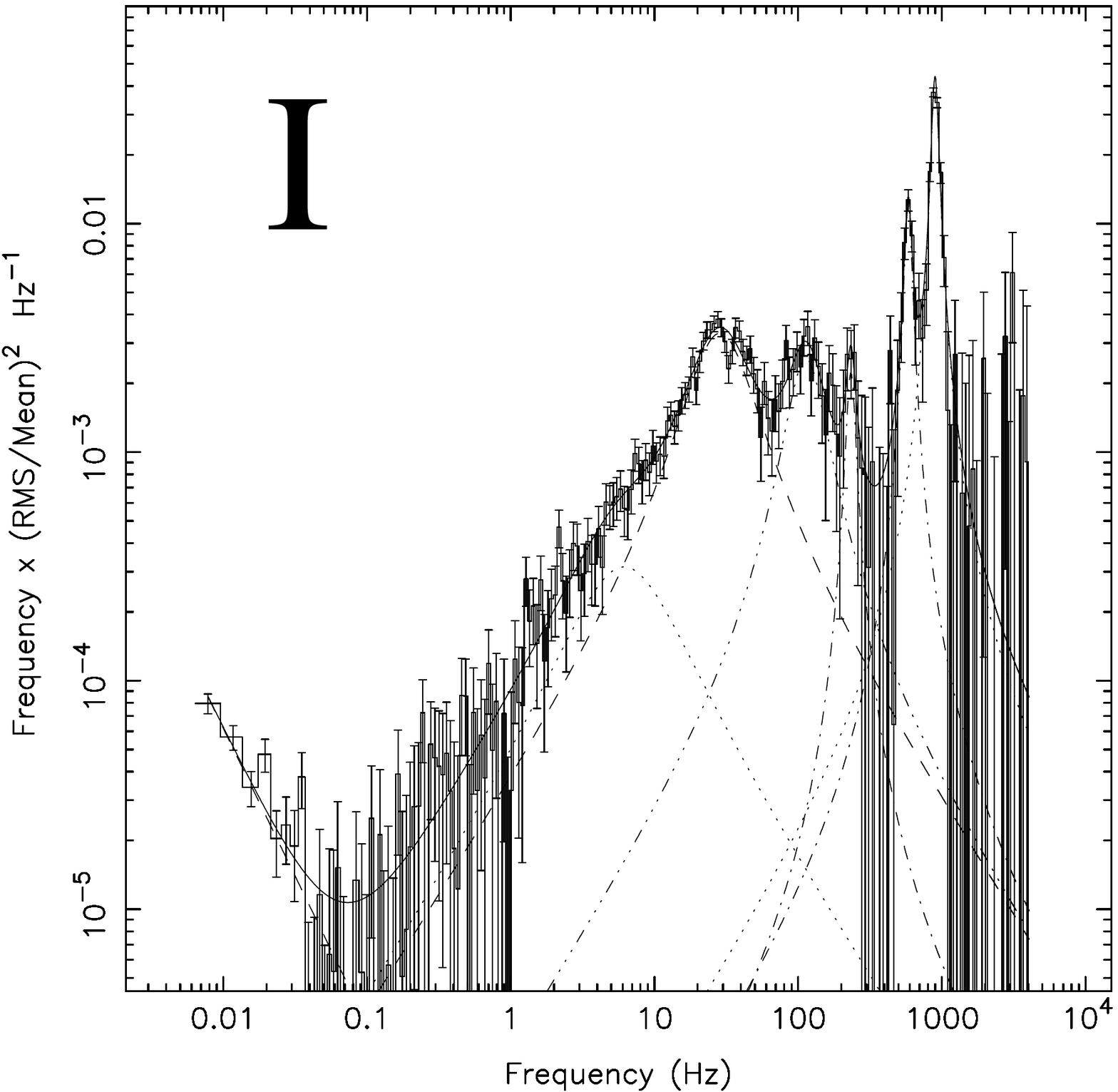}}}                     
\resizebox{0.4\columnwidth}{!}{\rotatebox{0}{\includegraphics{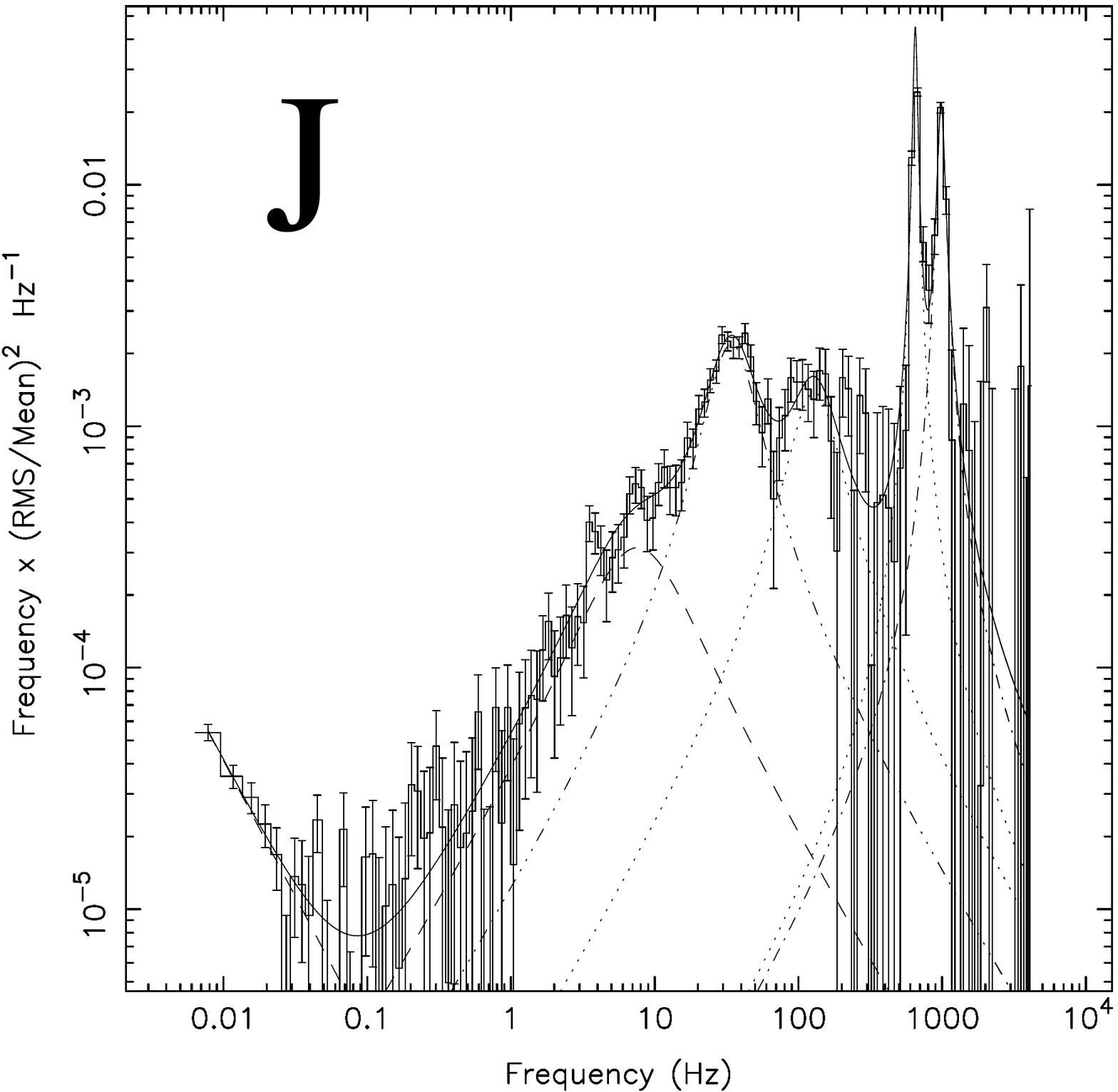}}}                
\resizebox{0.4\columnwidth}{!}{\rotatebox{0}{\includegraphics{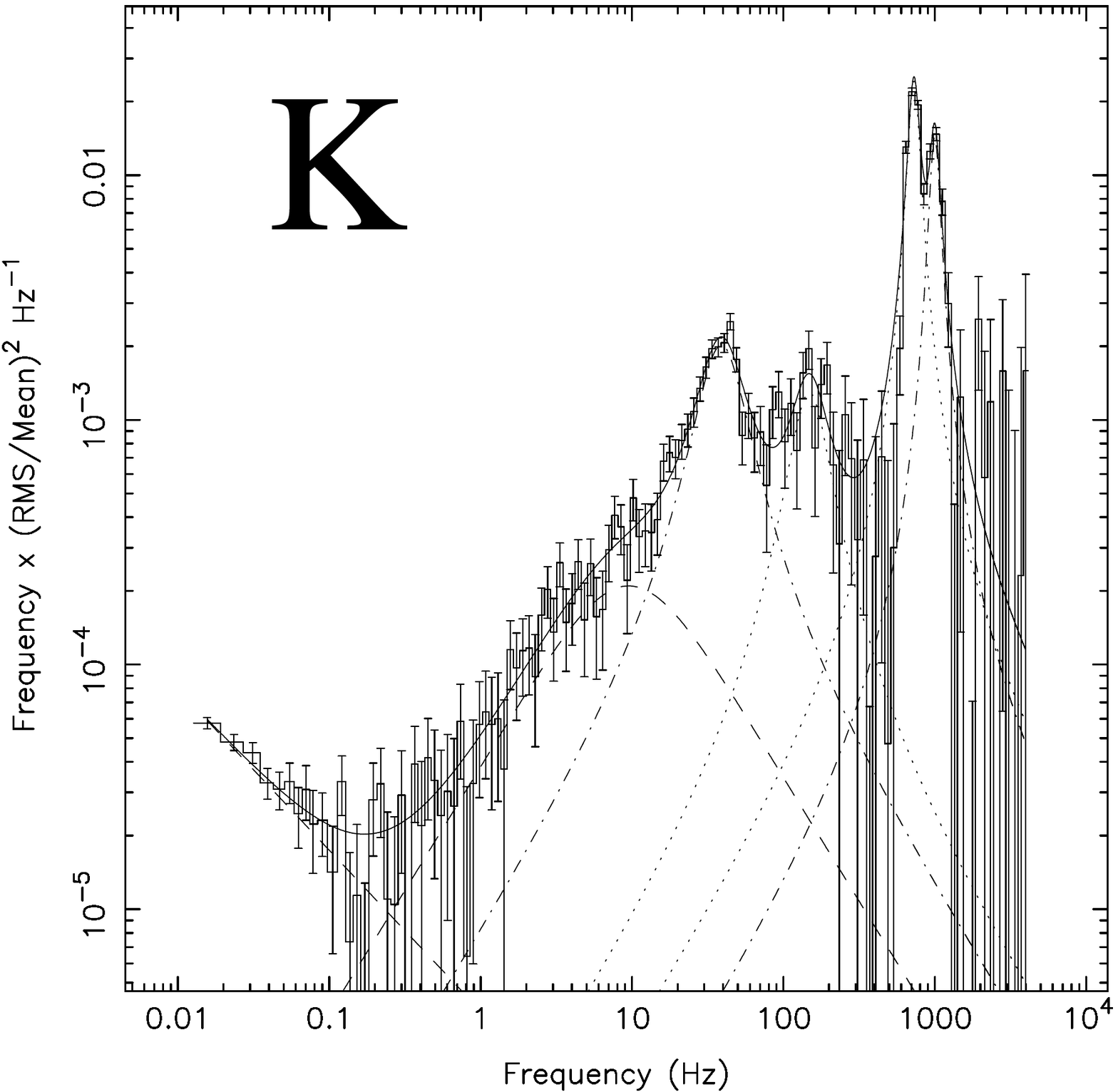}}}               
\resizebox{0.4\columnwidth}{!}{\rotatebox{0}{\includegraphics{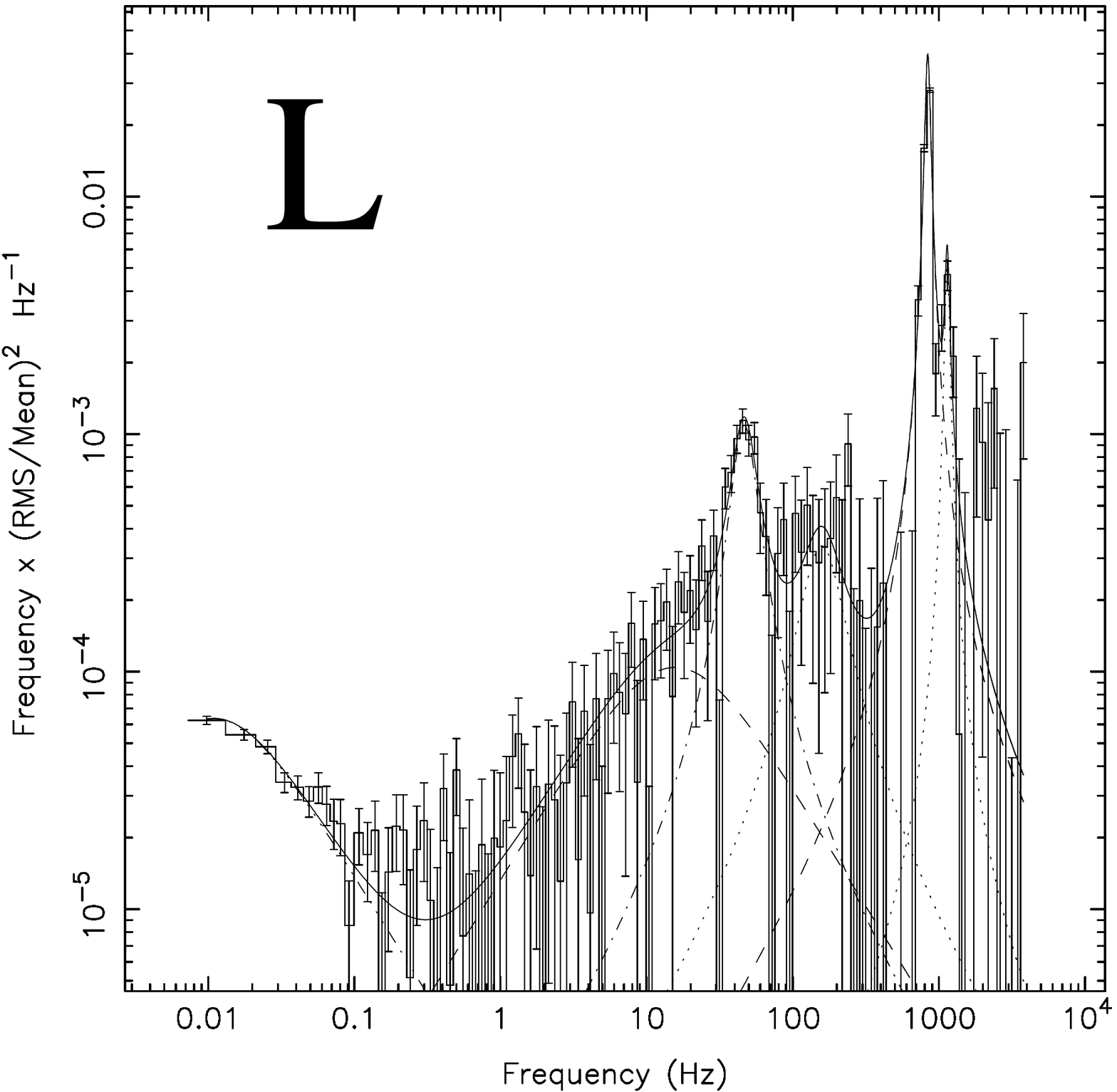}}} 
\resizebox{0.4\columnwidth}{!}{\rotatebox{0}{\includegraphics{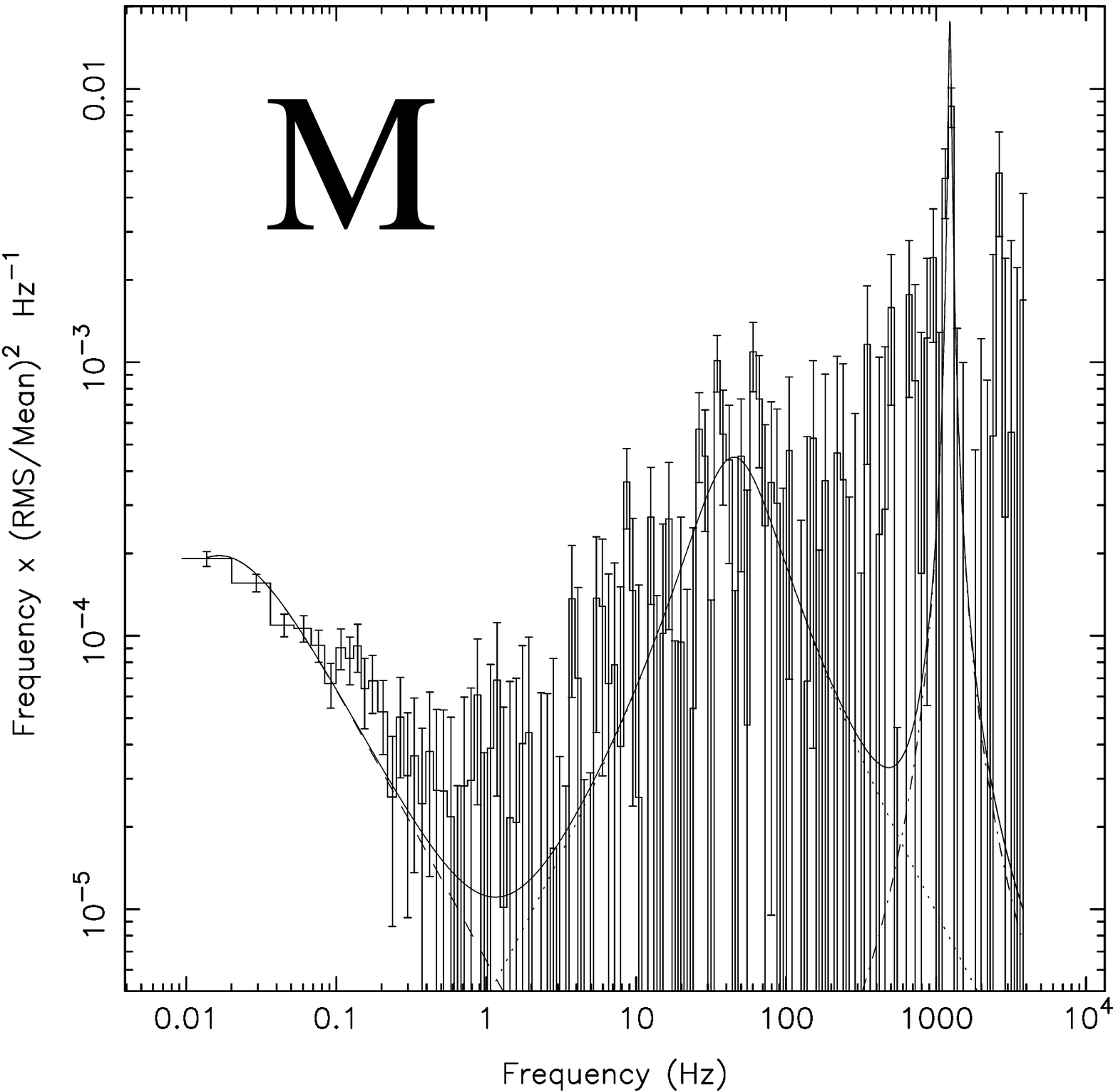}}}                     
\resizebox{0.4\columnwidth}{!}{\rotatebox{0}{\includegraphics{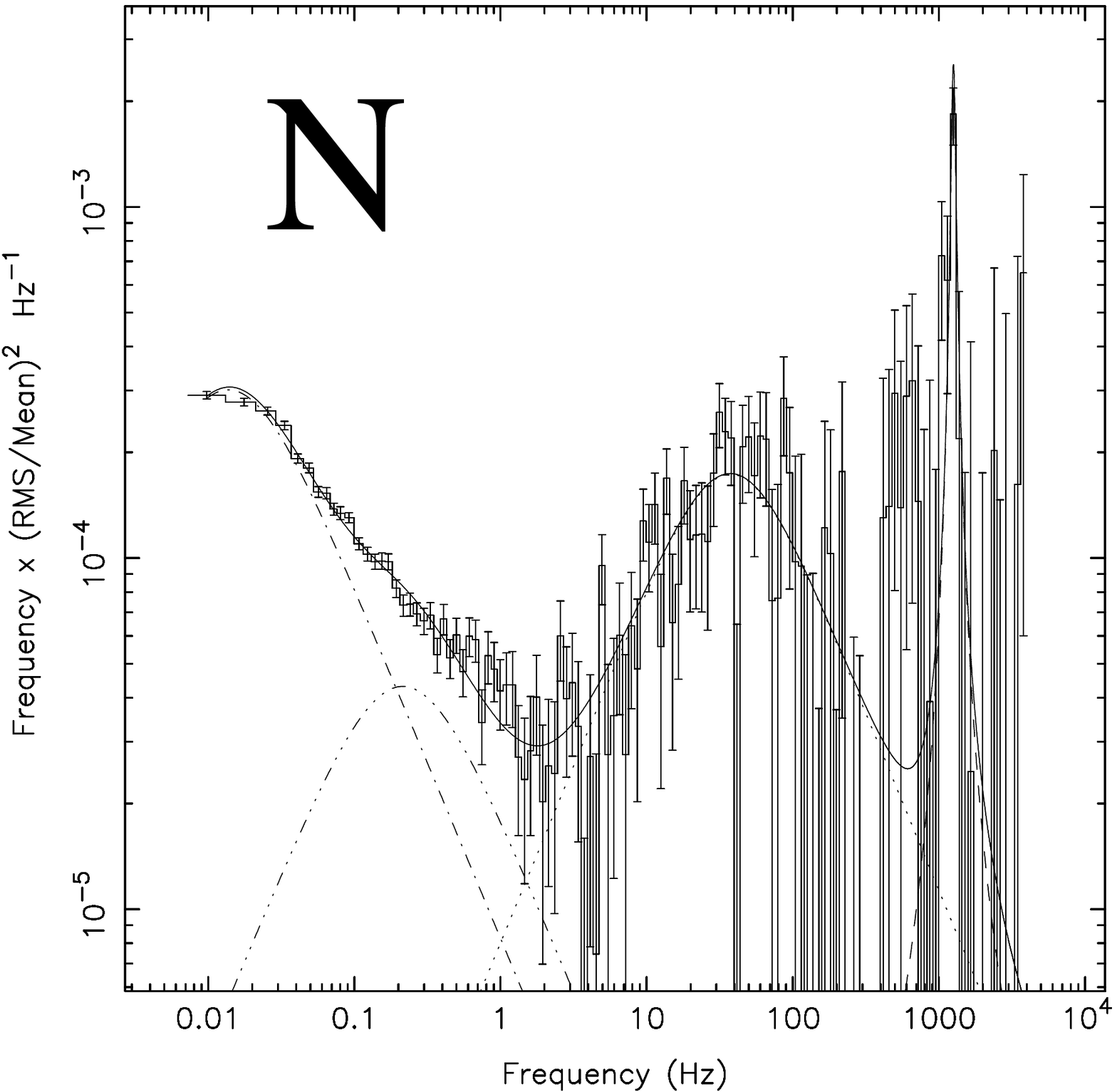}}}                     
\caption{Power spectra and fit functions in the power spectral density times frequency representation. 
Each plot corresponds to  a different region in the color-color
and color-intensity diagrams (See Figures \ref{fig:ccd} and \ref{fig:cvsint}). 
The curves mark the individual Lorentzian components of the fit. 
For a detailed identification, see Table~\ref{table:data} and Figure~\ref{fig:nuvsnu}.}
\label{fig:powerspectra}
\end{figure*}

\begin{figure*}[!hbtp] 
\center
\resizebox{0.80\columnwidth}{!}{\rotatebox{0}{\includegraphics{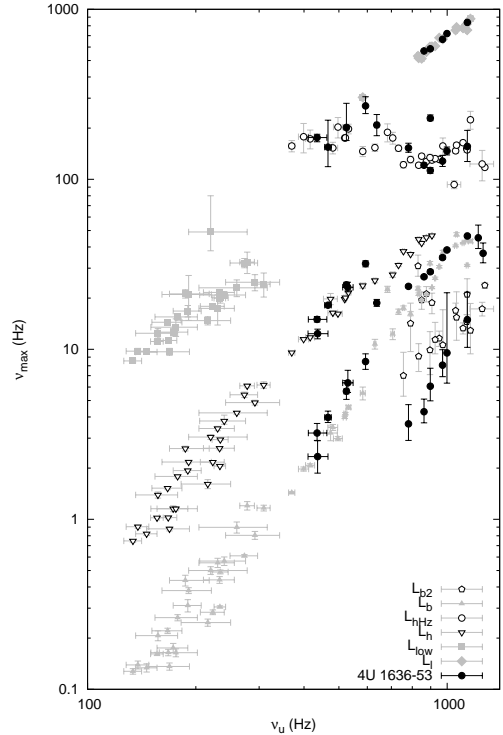}}}
\caption{The characteristic frequencies $\nu_{max}$ of the various power
spectral components plotted versus $\nu_{u}$. 
The black bullets mark the results for 4U~1636--53.
The other symbols mark the atoll sources  
4U~0614+09, 4U~1728--34 \citep{Straaten02}, 
4U~1608--52 \citep{Straaten03}, and Aql~X-1 \citep{Reig04}
and the low luminosity bursters 1E~1724--3045, GS~1826--24 and SLX
1735--269 \citep[][but also see \citealt{Belloni02}]{Straaten05}.}
\label{fig:nuvsnu}
\end{figure*}

In Figure~\ref{fig:powerspectra} we show the average power spectra
$A-N$ with their fits.  Two to five Lorentzian components were needed
for a good fit, except in power spectrum I, where an extra component
is needed, for a total of six.  Table~\ref{table:data} gives the fit
results. Power spectra A1 and A2 have the same $\nu_u$ within errors,
and only slightly different colors. We treat these two power spectra
separately because A1 could be fitted with 5 significant components
and A2, as well as the combined spectrum A1+A2, only with 3. Note,
that power spectrum A1 is the average of one observation (See
Table~\ref{table:averages}) which was performed in between the
observations used for power spectrum A2.
In Figure~\ref{fig:nuvsnu}
we show our measured characteristic frequency correlations (black)
together with those previously measured in other atoll sources (grey).
In intervals H--L, the twin kilohertz QPOs ($L_u$ and $L_\ell$) are
identified unambiguously.  The correlation between the lower and the
upper kilohertz QPO is the same as that found in the other atoll
sources studied by \citet{Straaten03}.  For intervals M and N, only
$L_u$ is observed.

\begin{figure}[!hbtp]
\centering 
\resizebox{0.50\columnwidth}{!}{\rotatebox{0}{\includegraphics{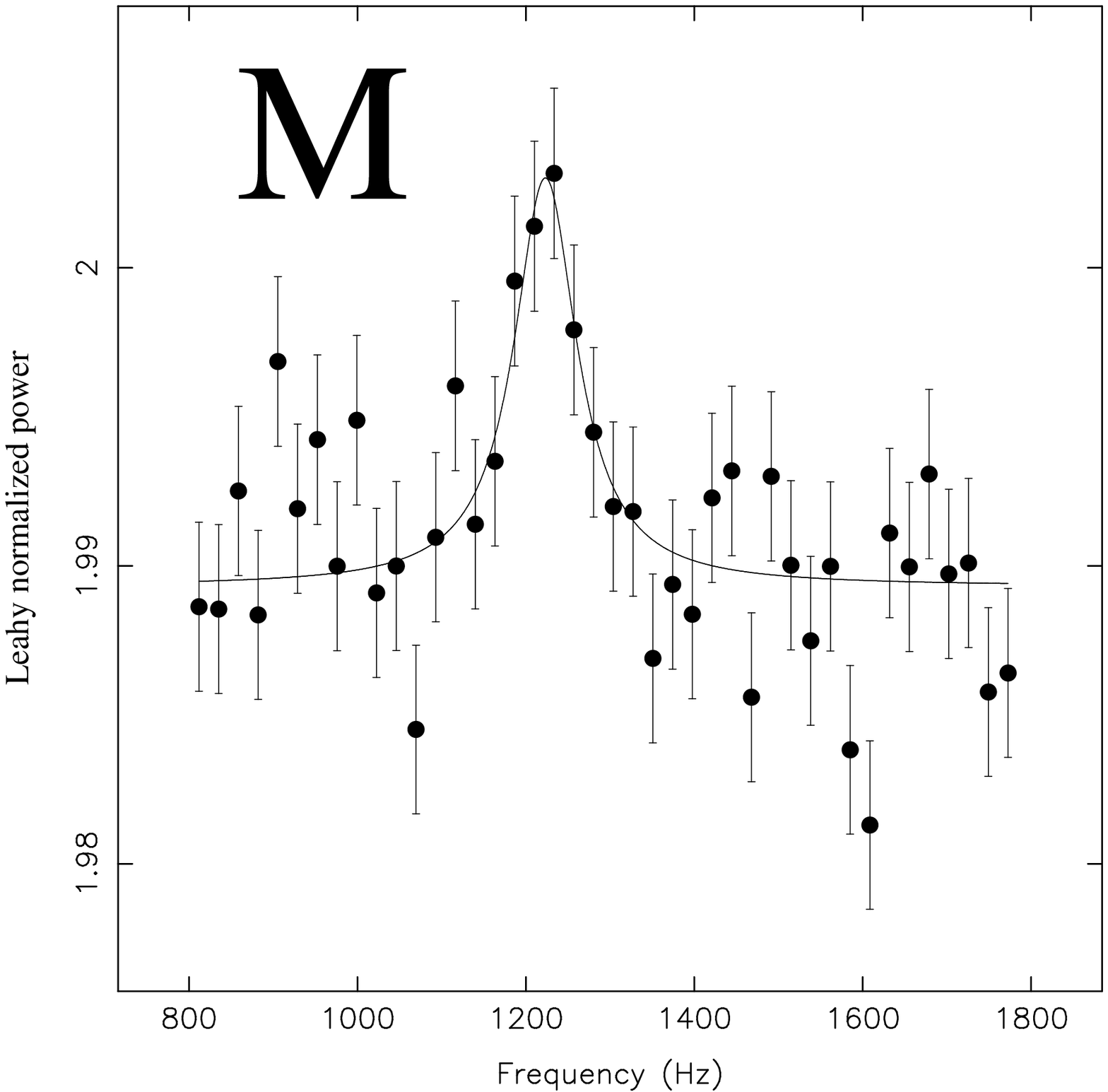}}}
\resizebox{0.50\columnwidth}{!}{\rotatebox{0}{\includegraphics{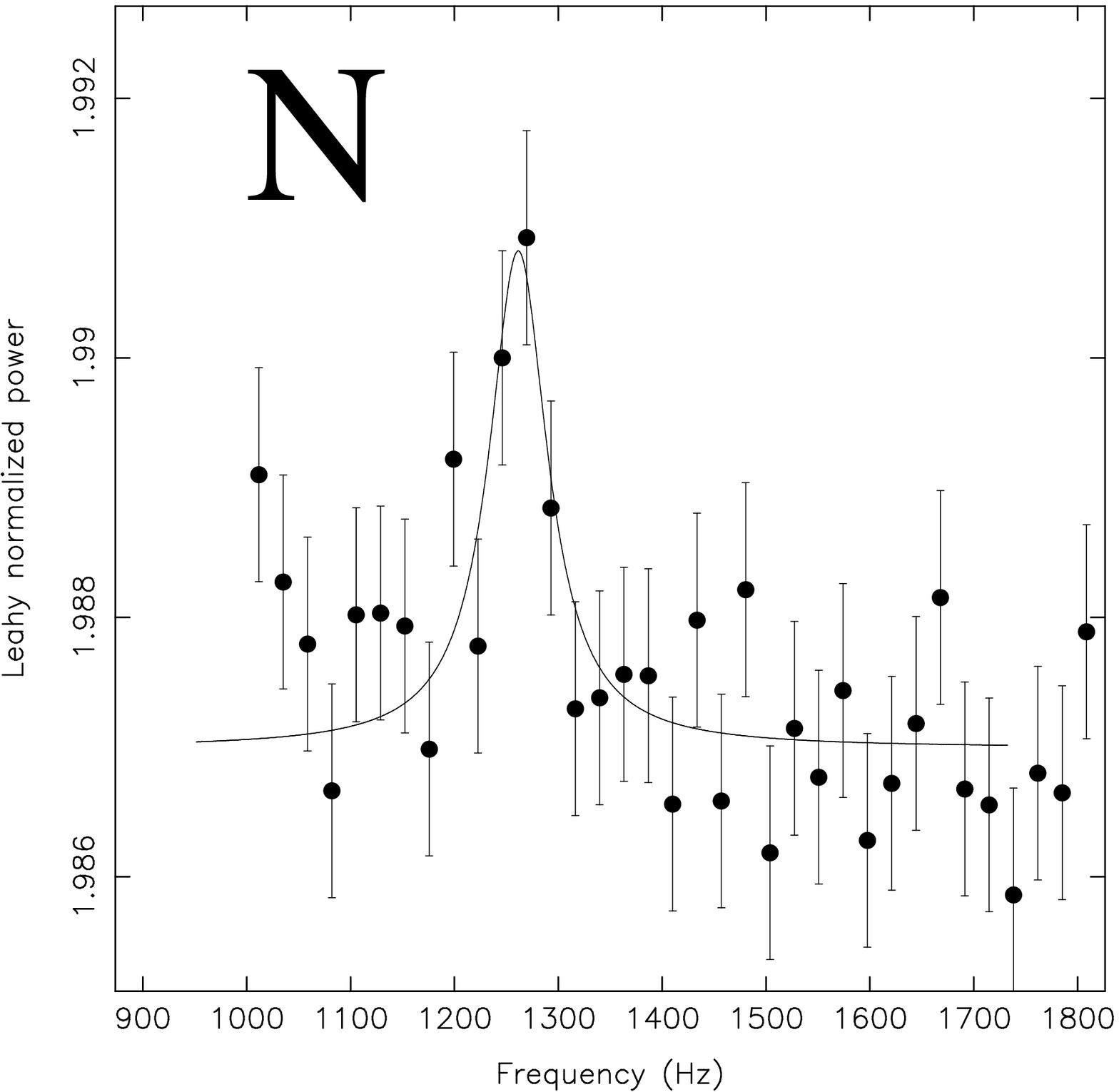}}}
\caption{kHz QPOs of intervals $M$ and $N$, respectively. 
These are Leahy normalized power spectra with no Poisson noise subtraction.}
\label{fig:mnzoom}
\end{figure}

We separately measured the centroid frequencies $\nu_0$ of the kHz
QPOs (see Table~\ref{table:centralfreq}).  The centroid frequency
difference $\Delta\nu$ varied between $274.6\pm5.7$~Hz (in interval K)
and $316.4\pm5.7$~Hz (interval I).  These values are between the
extremes found by \citet{Jonker02} ($\Delta\nu= 330\pm9$~Hz) and
\citet{Disalvo03} ($\Delta\nu=242\pm4$~Hz).  Note that those authors
used much shorter time intervals to detect the kHz QPOs and hence were
more sensitive to short lived extreme cases.  Our average power
spectra contain those data which is necessary to detect the broad
components well, and hence our measured $\Delta\nu$'s occur at
intermediate values, which is when the kHz QPOs are
strongest \citep[see e.g.][]{Mendez01}. So our method averages out the
extreme cases.  In power spectrum N, $L_u$ reaches the highest centroid
frequency found among the intervals analyzed, $1259\pm10$~Hz.  Note that
power spectra M and N are the result of averaging large amounts of
data with no significant kHz QPOs in individual observations
($\sim2.2\times10^{4}$ seconds and $\sim2.5\times10^5$ seconds,
respectively), based on the position of 4U~1636--53 in the color-color
diagram.  Figure~\ref{fig:mnzoom} displays the upper kHz QPOs in power
spectra M and N more clearly.

\begin{figure}[!hbtp] 
\center
\resizebox{0.9\columnwidth}{!}{\rotatebox{0}{\includegraphics{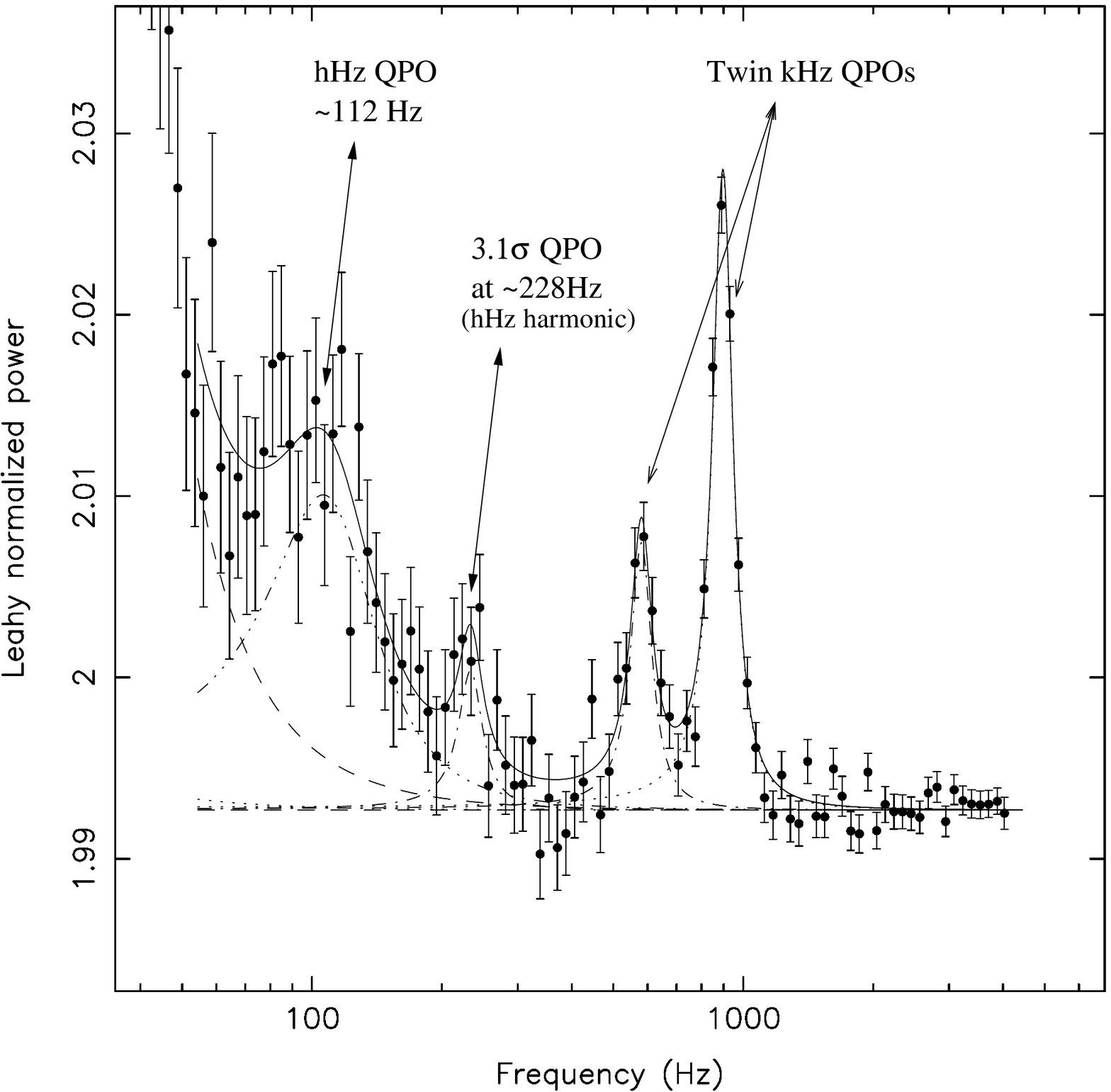}}}
\caption{Part of the power spectrum of interval I, showing the twin kHz QPOs, 
the hHz QPO and its possible harmonic. This is a Leahy normalized power spectrum 
with no Poisson noise subtraction.}
\label{fig:harmonic}
\end{figure}

Between $\sim100$~Hz and $\sim200$~Hz, a Lorentzian with quality
factor $Q\sim1$ is often found in atoll sources \citep[see][and
references therein]{Vanderklis04}. This feature is called hectohertz
QPO and, in contrast to the other components, its frequency remains
confined to this relatively narrow range as $\nu_u$ increases. When
$\nu_u \gtrsim 800$~Hz, the twin kilohertz QPOs are usually identified
unambiguously, and so is the hectohertz QPO.  For $\nu_u \lesssim 600$
Hz, the lower kHz QPO could also have frequencies between $\sim100$~Hz
and $\sim300$~Hz if it would be present, which makes it difficult to
classify the QPOs found in that range as either $L_{\ell}$, $L_{hHz}$
or a blend without more information.  In our data, this is the case
for intervals A to E; in Table~\ref{table:data} and hereafter we
identify those Lorentzians as hectohertz QPOs.  This identification is
supported by the fact that in intervals F and G, i.e., for 600
$<\nu_u<$ 800~Hz, $L_{\ell}$ is undetected; this component seems to
appear only at $\nu_u> 800$~Hz.  Interval I shows a $\sim3.1\sigma$
(single trial) peak with $\nu_{max} = 229\pm9$~Hz which is a factor
$\sim$2 higher than the usual hHz 
in that range. This QPO may be the second harmonic of the hectohertz
QPO simultaneously found at a characteristic frequency of $113\pm4$~Hz
(see Figure~\ref{fig:harmonic}, however, note that this feature must
be interpreted with care due to its low -3.1$\sigma$- single trial
significance).  As a result of refitting the power spectrum using
centroid frequencies, we find that the second harmonic QPO is at
$\nu_0 = 228 \pm 10$~Hz while the first harmonic hHz QPO is at $\nu_0
= 107 \pm 5$~Hz for a ratio of $2.13 \pm 0.13$, consistent with 2.

\begin{figure}[!hbtp]
\centering 
\resizebox{1\columnwidth}{!}{\rotatebox{-90}{\includegraphics{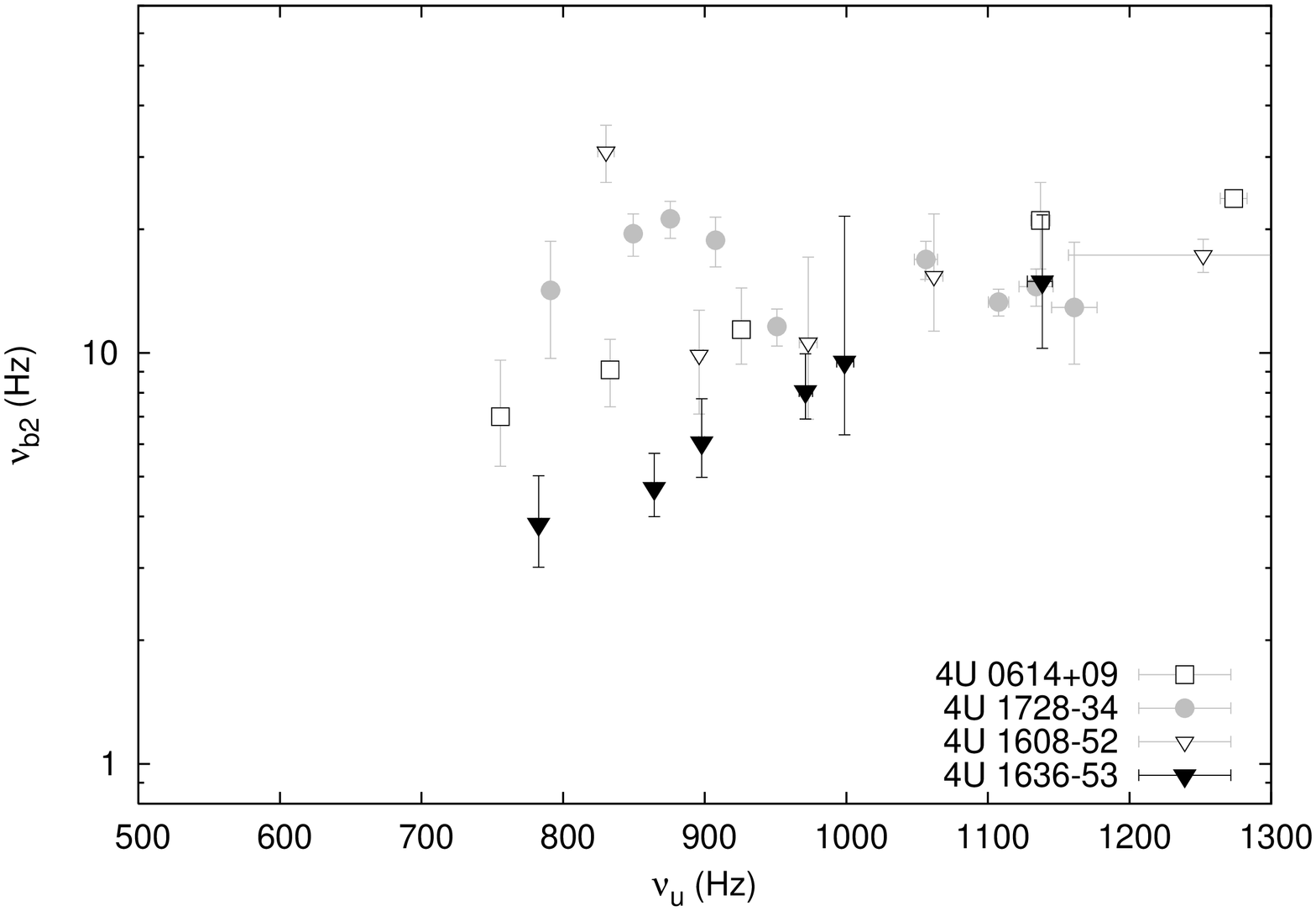}}}
\caption{$\nu_{b2}$ versus $\nu_u$ for the 4 atoll sources 
4U~0614+09, 4U~1728--34 \citep{Straaten02}, 
4U~1608--52 \citep{Straaten03} and 4U~1636--53 (this paper).
Note that the open square at $\nu_u\sim1233$~Hz 
 could be interpreted as either $L_b$ or $L_{b2}$ \citep{Straaten02}.
}
\label{fig:b2}
\end{figure}

Figure~\ref{fig:nuvsnu} shows that also $L_h$ and $L_b$ lie on the
correlations previously observed in other sources.  However, our
results show that $\nu_b$ may anti-correlate with $\nu_u$ at
$\nu_u\gtrsim1100$~Hz. This result has already been observed in Z
sources, however for atoll sources this behavior has not been observed
with certainty (See Section~\ref{sec:discussion}).  $L_{b2}$ seems to
have lower frequencies than in other atoll sources. To further
investigate this, in Figure~\ref{fig:b2} we plot $\nu_{b2}$ versus
$\nu_u$ with different symbols for each of the 4 atoll sources for
which this component has been measured. Clearly, the range in which
$L_{b2}$ has been found for similar $\nu_u$ is rather large (up to
nearly a decade), particularly at $\nu_u < 1000$~Hz.
 We also studied the possibility that the rms of $L_{b2}$ could be
related with its frequency, but no relation was found.


\begin{figure}[!hbtp] 
\center
\resizebox{0.8\columnwidth}{!}{\rotatebox{-90}{\includegraphics{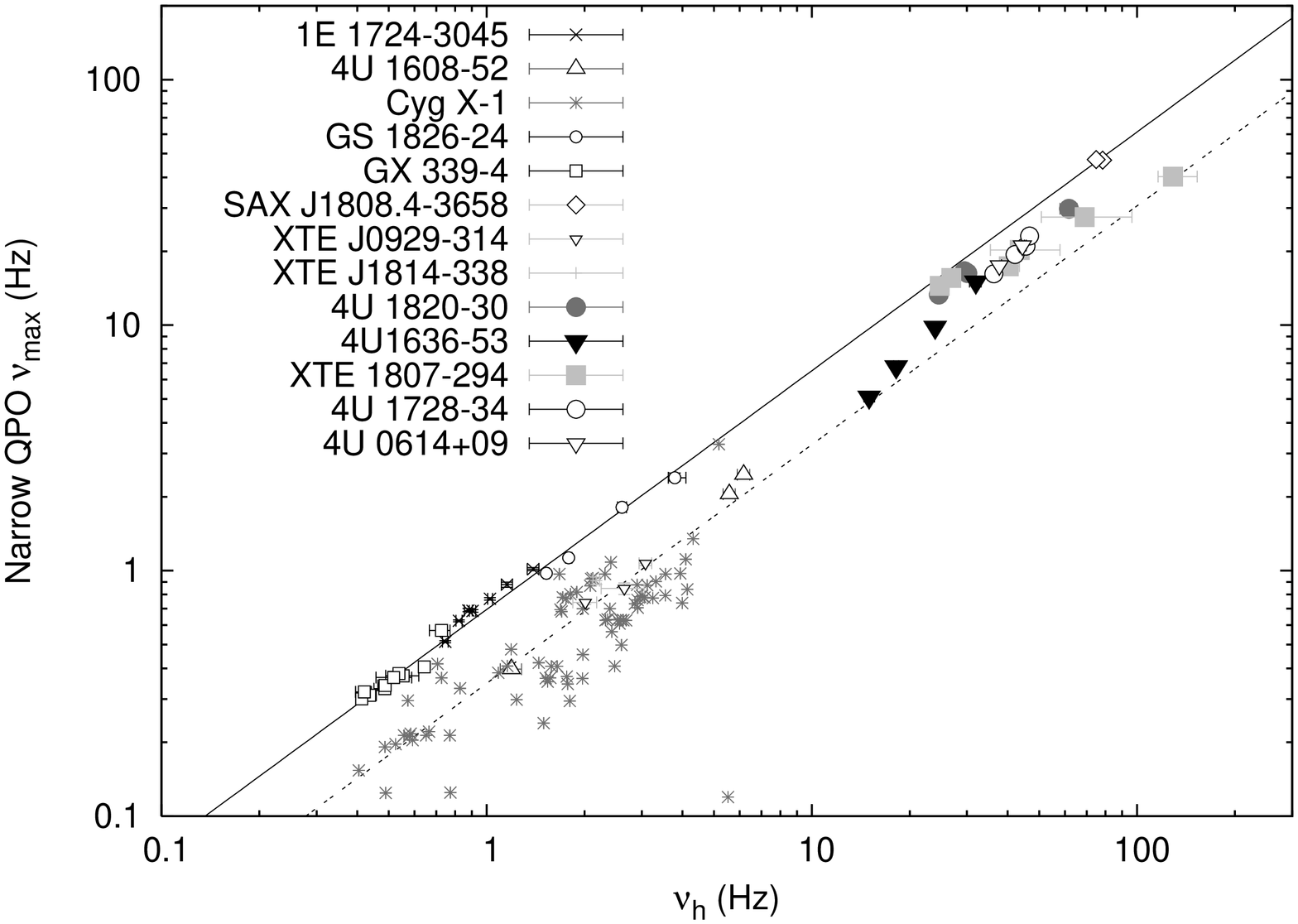}}}
\resizebox{0.8\columnwidth}{!}{\rotatebox{-90}{\includegraphics{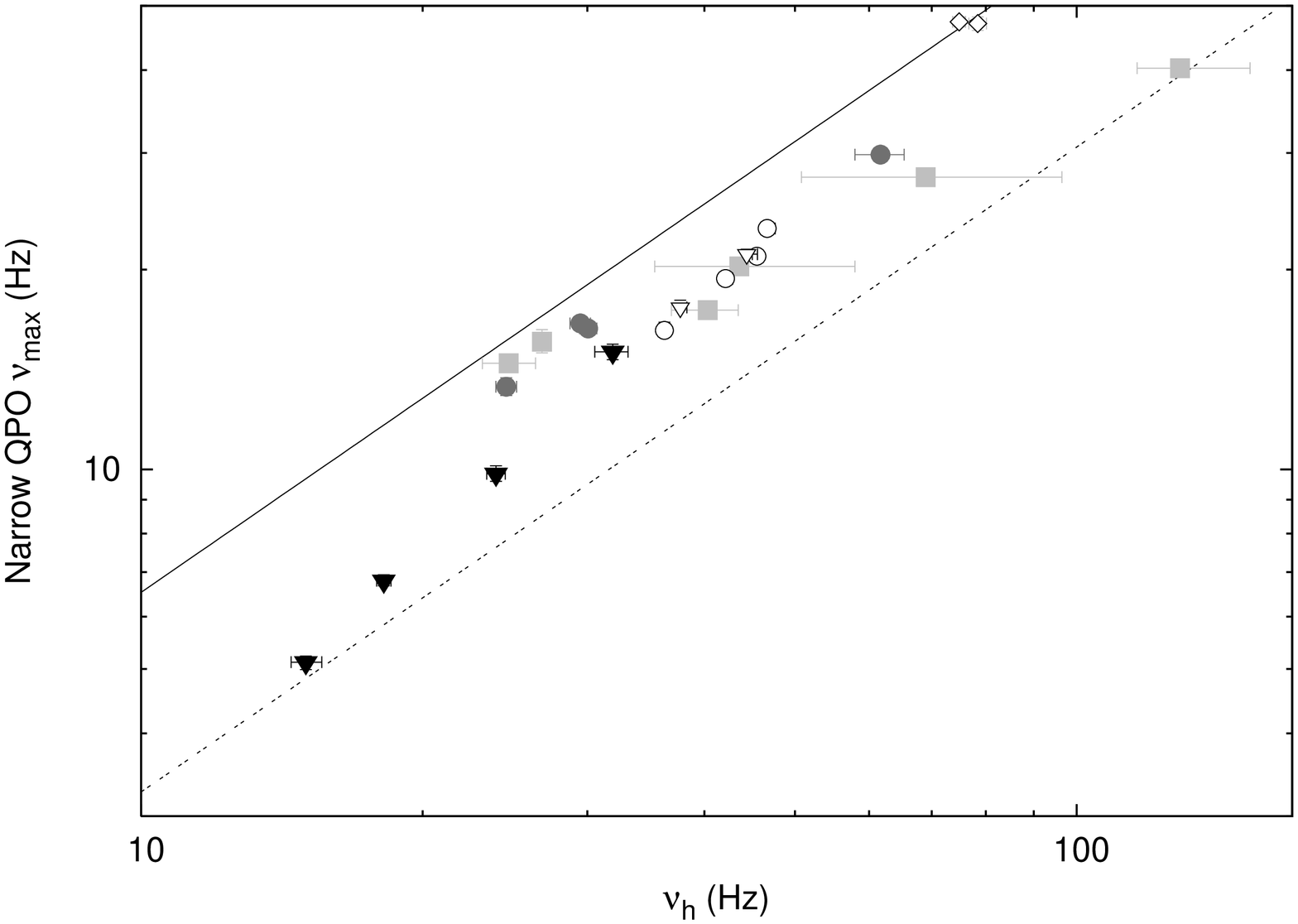}}}
\caption{ \textit{(Above)} Characteristic frequencies  $\nu_{LF}$ and $\nu_{LF/2}$ (see
text)  versus $\nu_h$  .  The  symbols are  labeled in  the  plot, and
represent  the   frequencies  of  the   QPOs  in  the   atoll  sources
4U~1728--34,  4U~0614+09, 4U~1608--52, 4U~1820--30,  the BHCs  Cyg X-1
and   GX~339--4,  the  low   luminosity  bursters   1E~1724--3045  and
GS~1826--24  and  the  accreting millisecond  pulsars  XTE~J0929--314,
XTE~J1814--338                  and                  SAX~J1808.4--3658
\citep{Straaten03,Straaten05,Altamirano05}. The  filled triangles show
the results for 4U~1636--53.  The drawn line indicates a power-law fit
to the $\nu_{LF}$ vs.  $\nu_h$ relation of the low-luminosity bursters
1E~1724--3045 and GS~1826--24, and the BHC GX~339--4.  The dashed line
is  a  power law  with  a  normalization half  of  that  of the  drawn
line. \textit{(Bottom)} A zoom of the high frequency region. }
\label{fig:bumps}
\end{figure}

Intervals A, B, D and E show a narrow QPO with a characteristic
frequency between $\nu_b$ and $\nu_h$ (see Table~\ref{table:data}).
For other neutron stars such narrow QPOs were previously reported by
\citet{Yoshida93} in 4U~1608--52, by \citet{Belloni02} in the low
luminosity bursters 1E~1724--3045 and GS~1826--24, by
\citet{Straaten02} in 4U~0614+09, 4U~1728--34, by \citet{Straaten03}
in 4U~1608-52, by \citet{Altamirano05} in 4U~1820--30, by
\citet{Straaten05} in the accreting millisecond pulsars (AMP)
XTE~J0929--314, XTE~J1814--338 and SAX~J1808.4--3658 and by
\citet{Manu05} in XTE~J1807--294.  Similar features were also seen in
the BHCs Cyg~X-1 by \citet{Pottschmidt03} and GX~339--4 by
\citet[][but also see \citealt{Straaten03}]{Belloni02}.
Following \citet{Straaten03}, for clarity we have omitted these QPOs
($L_{LF}$) from Figure~\ref{fig:nuvsnu}. In Figure~\ref{fig:bumps} we
plot their characteristic frequencies vs. $\nu_h$. The results for
4U~1636--53 are in the range of, but seem to follow a different
relation than, the two relations previously suggested by
\citet{Straaten03} based on other sources. 
The $L_{LF}$ QPOs in 4U~1636--53 cannot be significantly
detected on  timescales shorter than the duration of  an average observation.

\begin{figure}[!hbtp] 
\center
\resizebox{0.7\columnwidth}{!}{\rotatebox{-90}{\includegraphics{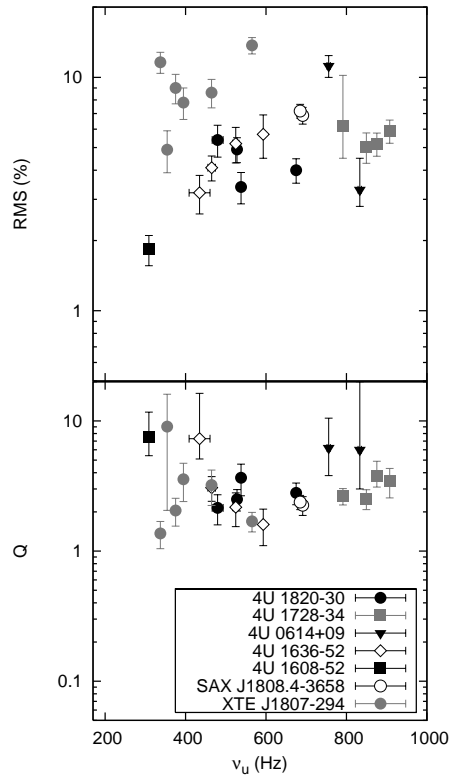}}}
\caption{$L_{LF}$'s rms (\textit{above}) and Q (\textit{below}) versus
$\nu_u$ for 7 atoll sources. The symbols are labeled are the same in
both plots. The scatter of the points are likely due to the averaging
method (see text).}

\label{fig:LF}
\end{figure}

For completeness, in Figure~\ref{fig:LF} we plot both the
fractional rms amplitude and quality factor of the $L_{LF}$ component
versus $\nu_u$.  Although the fractional rms amplitude of 4U~1636--53
increases with $\nu_u$, no general trend is observed among the 7
sources shown in this Figure.  The quality factor $Q_{LF}$ (which may
have been affected by smearing, see
Section~\ref{sec:dataanalysis}) seems to be unrelated to $\nu_u$ for
all the sources shown.

\begin{figure*}[!hbtp] 
\center
\resizebox{2\columnwidth}{!}{\rotatebox{-90}{\includegraphics{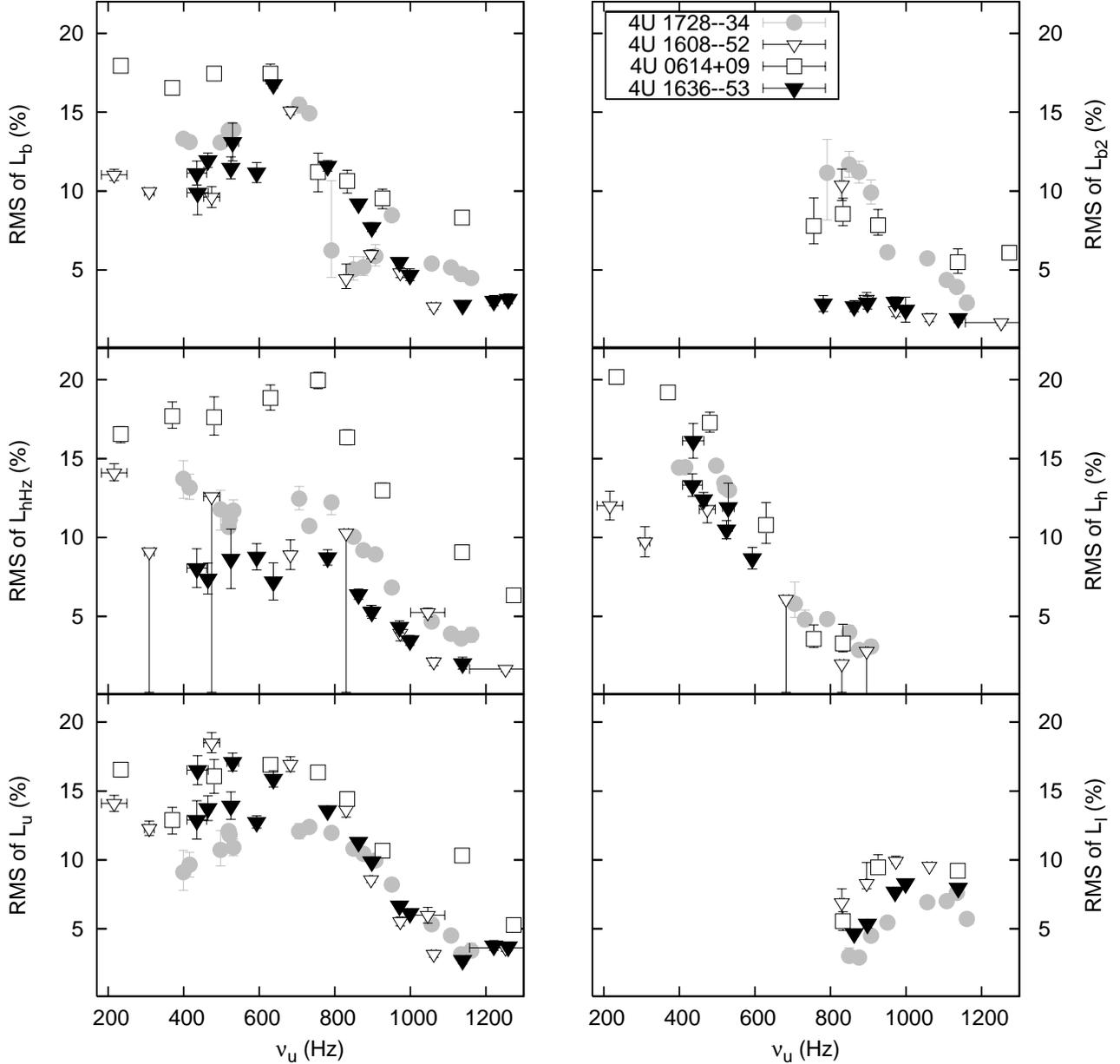}}}
\caption{The fractional rms amplitude of all components (except $L_{LF}$) plotted versus $\nu_{u}$. 
The symbols are labeled in the plot. The data for 4U~1728--34, 4U~1608--52 and 4U~0614+09
were taken from \citet{Straaten05}. Note that for $L_{hHz}$ and $L_{h}$ of 4U~1608--52, the 
3 triangles with vertical error bars which intersect the abscissa represent 95\% confidence upper limits
(see \citet{Straaten03} for a discussion). Also note that we exclude the 3 points in the $rms_{\ell}$ versus $\nu_u$ plot which were identified as $L_{low}$ by \citet{Straaten03}. }
\label{fig:rmsvsnu}
\end{figure*}

In Figure~\ref{fig:rmsvsnu} we plot the fractional rms amplitude of
all components (except $L_{LF}$, see Figure~\ref{fig:LF}) versus
$\nu_{u}$ for the atoll sources 4U~0614+09, 4U~1728--34, 4U~1608--52
and 4U~1636--53.  The rms of the upper kHz QPO for all sources
approximately follows the same trend: it increases up to
$\nu_u\sim750-800$~Hz, and then starts to decrease. This seems also to
be the case for $L_{hHz}$ and $L_b$.  Except for 4U~0614+09, the data
suggest that at $\nu_u\gtrsim1100$~Hz the rms of $L_u$ does not
decrease further, but remains approximately constant.
At $\nu_u\sim750-800$~Hz, the rms amplitudes of $L_{hHz}$
and $L_b$ start to decrease \citep[see also][]{Straaten03}.
The rms of $L_h$ of 4U~1636--53 also seems to follow the general
trends observed for the other atoll sources. Some of these results
were previously reported by \citet{Straaten03}, \citet{Mendez01} and
\citet{Barret05}. 4U~1636--53 stands out by the fact that the rms of
$L_{b2}$ and $L_{hHz}$ at $\nu_u \lesssim 900$~Hz is always smaller
than in the other sources. Moreover, and again contrary to the other
sources, in 4U~1636--53 the rms of $L_{b2}$ remains approximately
constant as $\nu_u$ increases from $\sim800$~Hz to $\sim1200$~Hz.

\begin{figure}[!hbtp] 
\center
\resizebox{0.8\columnwidth}{!}{\rotatebox{-90}{\includegraphics{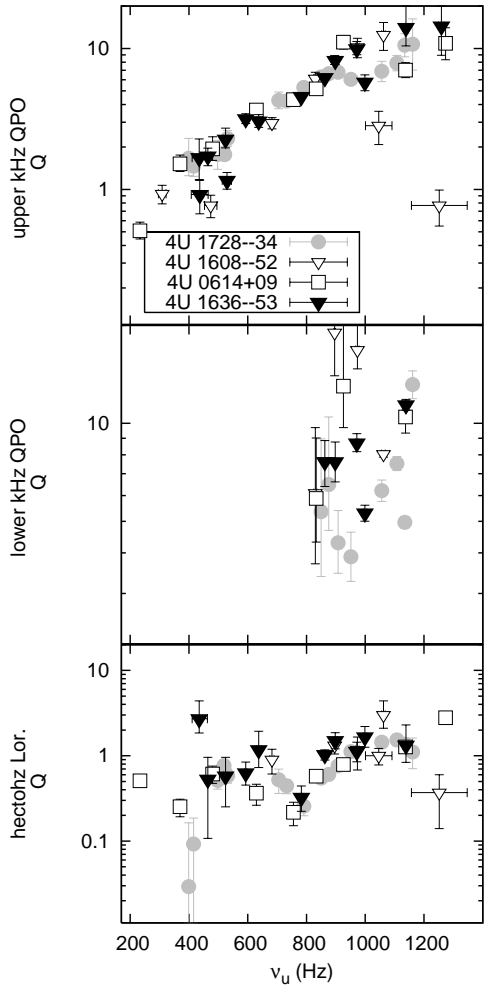}}}
\caption{Quality factor Q of $L_u$, $L_{\ell}$ and $L_{hHz}$ versus
$\nu_u$. Symbols are the same as in Figure~\ref{fig:rmsvsnu}. Note
that the results for the quality factor are probably affected by the
averaging method (See Sections~\ref{sec:intro}, \ref{sec:results} and
the Appendix). }
\label{fig:nuvsq}
\end{figure}

In Figure~\ref{fig:nuvsq} we plot the quality factors $Q_{u}$,
$Q_{\ell}$ and $Q_{hHz}$ versus $\nu_{u}$.  As noted in
Section~\ref{sec:dataanalysis}, the Q values of $L_{\ell}$, and to a
lesser extent, $L_u$, have likely been affected by smearing.  The Q
values of the other components are not plotted since they are usually
broad.  The data on 4U~1636--53 are in general agreement with what was
found using similar methods on the other sources \citep{Straaten02}.
$Q_u$ increases monotonically with $\nu_u$ until
$\nu_u\sim900-1000$~Hz.  At this frequency, $Q_u$ seems to decrease
for all sources, to immediately increase again as $\nu_u$ increases.
$Q_{\ell}$ shows a rather random behavior due to smearing (see
\citealt{Disalvo03} and \citealt{Barret05,Barret05a} for $Q_{\ell}$
measurements less affected by smearing); we display this quantity in
Figure~\ref{fig:nuvsq} for comparison to previous works using the same
method.

$Q_{hHz}$ shows a complicated behavior.  To
further investigate this behavior, we first re-binned the $Q_{hHz}$
data by a factor 3 and fitted a straight line. The best fit gives
$\chi^2/dof=93 / 12 \sim7.7 $. Since the data appears to show two
bumps separated by a minimum creating a roughly sinusoidal pattern, we
also tried to fit a straight line plus a sine wave. The best fit has a
$\chi^2/dof=16.9 / 9 \sim1.8$ which gives a $3.4\sigma$ improvement of
the fit based on an F-test. 
In Figure~\ref{fig:nuvsnu} it can be seen that $\nu_{hHz}$ appears to
display a similar pattern.  Fitting the relation of $\nu_{hHz}$ versus
$\nu_u$ with only a straight line gives a $\chi^2/dof=156 / 12 =13$
while a straight line plus a sine gives a $\chi^2/dof=28.4 / 9
\sim3.15$, once again, we find a $3.4\sigma$ improvement of the fit
based on an F-test.  The results of the sinusoidal fits, in which the
parameter errors have been rescaled by the reduced $\sqrt{\chi^2}$ value,
are given in Table~\ref{table:fit}.

With the present data it is difficult to distinguish if this is the
behavior of $L_{hHz}$ alone, or  is due to blending with other
components which are not strong and coherent enough to be observed
separately on short time scales and are lost in the averaging of the
power spectra.
This, as well as other ambiguities (see discussion about the
identification of $L_{\ell ow}$ in \citealt{Straaten03}), arise because of
the gaps between the $\nu_{\ell}$, $\nu_{\ell ow}$, $\nu_h$ and
$\nu_{hHz}$ versus $\nu_u$ relations (See Figure~\ref{fig:nuvsnu}).
Although we are not able to explain those gaps, the interpretation
that $L_{hHz}$ could be affected by the presence of other components
is made more likely by the fact that $L_{\ell}$ in Z sources can be
unambiguously identified down to frequencies of $\sim150$~Hz (see
e.g. \citealt{Jonker02b}) and that $L_h$ and $L_{\ell ow}$ have
sometimes been observed at frequencies up to $\sim120$ and
$\sim60$~Hz, respectively, i.e., in both cases reaching the hectohertz QPO
range (\citealt{Straaten05}, \citealt{Manu05}).  If it would be
possible to follow a source in its evolution from the extreme island
states where $L_{\ell ow}$ is prominent, to the lower left banana
where $L_{\ell}$ is seen, then some of these ambiguities could be
resolved.

\begin{table}[!hbtp]
\center
\begin{tabular}{|c|c|c|}\hline 
\hline
              & $Q_{hHz}$ & $\nu_{hHz} (Hz)$ \\
\hline
Slope          & $(1.6 \pm 0.2) \cdot 10^{-3} $ & $ (-5.7\pm 1.1) \cdot 10^{-2} $ \\
Constant       & $-0.52 \pm 0.15 $                & $196 \pm 10 $ \\
Amplitude      & $ 0.35\pm0.04$                   & $23.5 \pm 2.1 $ \\
Period         & $532\pm48$                       & $498\pm25$ \\
$\chi^2/dof$   &      15.1 / 15                    & 14.9 / 15 \\
\hline
\hline
\end{tabular}
\caption{The results of fitting the data of $Q_{hHz}$ and $\nu_{hHz}$
versus that of $\nu_u$ for the four atoll sources 4U~1728--34,
4U~0614+09, 4U~1608--52 and 4U~1636--53. We used the combination of a
straight line plus a sine.  The parameter errors have been rescaled
by the reduced $\sqrt{\chi^2}$ value (see text). The quoted errors use
$\Delta\chi^2 = 1.0$. }
\label{table:fit}
\end{table}

\section{Discussion}\label{sec:discussion}

In this paper, we report a detailed study of the time 
variability of the atoll source 4U~1636--53
using RXTE that includes, for the first time, observations of this
source in the (low-luminosity) island state.
We divided the data into 15 intervals, A to N, based on the position
of the source in the color diagram.  Based on the fact that, (i)
intervals A1, B, D and E show a narrow QPO with a characteristic
frequency between $\nu_b$ and $\nu_h$ which was previously seen in
other atoll sources when they were in their island state
\citep[e.g.][]{Straaten02,Altamirano05}; (ii) intervals A1, A2, B, D,
E and F do not show either $L_{\ell}$ or power-law VLFN at frequencies
lower than $0.5$~Hz, which would be expected to be present in the
banana state \citep{Hasinger89,Vanderklis04,Vanderklis06}; (iii) the intensity of
the source starts to increase from interval G (see
Figure~\ref{fig:cvsint}) and (iv) intervals A1 to F occupy the hardest
loci in the color diagram (see Figures \ref{fig:nuvsnu} and
\ref{fig:cvsint}), we conclude that intervals A1 to F show the source
in the island state, representing the first RXTE observations of
4U~1636--53 in this state.  Interval G may represent the transition
between the IS and the LLB since its power spectrum is very similar to
that of the first five intervals, but with the difference that a weak
($\sim1.2~\%$ rms) VLFN is present at a frequency lower than $0.1$~Hz
(see Figure~\ref{fig:nuvsnu}).

Along the color color diagram we find all seven power spectral
components that were already seen in other sources in previous works
\citep[see][for a review]{Vanderklis04}: $L_u$ is detected in all of
our power spectra, $L_{\ell}$ is unambiguously detected starting from
power spectrum H ($\nu_u \sim 800$~Hz), $L_{hHz}$ is observed in 11
out of 15 power spectra at frequencies between $100$ and $270$~Hz,
$L_h$ and $L_{LF}$ are detected mainly in the island state,
$L_b$ is always observed and finally
$L_{b2}$ is detected when $L_b$ becomes peaked from interval G.

Previous works have shown that the frequencies of the variability
components observed in other atoll sources follow a universal scheme
of correlations when plotted versus $\nu_u$ \citep[][and references
therein]{Straaten03}.  We have found that the noise and QPO
frequencies of the time variability of 4U~1636--53 follow similar
correlations as well (see Section~\ref{sec:results}) confirming the
predictive value of this universal scheme.  However, we also found
some differences between 4U~1636--53 and other atoll sources which we
discuss below. As 4U~1636--53 is one of the most luminous atoll
sources showing the full complement of island and banana states (full
atoll track), the object is of interest in order to investigate
the luminosity dependence of spectral/timing state behavior. This is of
particular importance to the ongoing effort to understand the origin
of the difference between the atoll sources and the more luminous Z
sources \citep[see e.g.][]{Homan06}.
\begin{figure}[!hbtp] 
\center
\resizebox{1\columnwidth}{!}{\rotatebox{-90}{\includegraphics{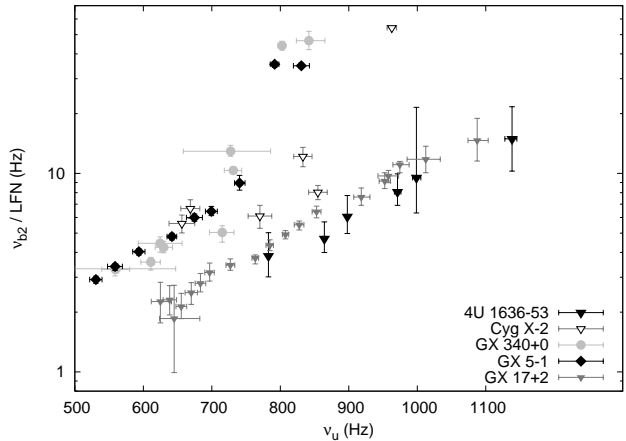}}}
\caption{$\nu_{b2}$ versus $\nu_u$ for the atoll source 4U~1636--53
(this paper) and the characteristic frequencies of the low-frequency
noise (LFN) versus $\nu_u$ for the Z sources
GX~17+2 \citep{Homan02}, 
Cyg X-2 \citep{Kuznetsov02}, 
GX~340+0 \citep{Jonker00} 
and GX~5--1 \citep{Jonker02} -- 
see text and \citet{Straaten03} for details.
(See also Figure~\ref{fig:b2})}
\label{fig:b2z}
\end{figure}

\subsection{The broad components in 4U~1636--53 and  Z-source LFN}
As can be clearly seen in Figure~\ref{fig:b2}, where we show
$\nu_{b2}$ versus $\nu_u$, the behavior of the $L_{b2}$ component
differs significantly between sources.  For 4U~1636--53 and
4U~0614+09, $\nu_{b2}$ increases with $\nu_u$, while this is not seen
for 4U~1608--52 and 4U~1728--34.  
This frequency behavior is different from that observed for
all other components (see Section~\ref{sec:results}), which instead is
very consistent between sources, even for the case of the hHz QPO,
which has not been seen to correlate with other components
\citep[see][for a review]{Vanderklis04}. This unusual, somewhat
erratic behavior of $L_{b2}$ may be related to the fact that it is
usually detected as a relatively weak wing to a much stronger $L_b$,
so that small deviations in the time-averaged shape of $L_b$ have a
large effect on $L_{b2}$.

In order to investigate the relation of $L_{b2}$ to the well-known low
frequency noise (LFN) which occurs in the same frequency range in Z
sources, in Figure~\ref{fig:b2z} we plot the results for 4U~1636--53
together with those for the LFN in the Z sources \citep{Hasinger89}
GX~17+2 \citep{Homan02}, Cyg X-2 \citep{Kuznetsov02}, GX~340+0
\citep{Jonker00} and GX~5--1 \citep{Jonker02b}.  Note that the
broad-band noise in these Z sources was not fitted with a
zero-centered Lorentzian but with a cutoff power law or a smooth
broken power law. We used the results of the conversion from power
laws to zero-centered Lorentzians done by \citet{Straaten03}.
Previous works \citep[e.g.][]{Psaltis99,Straaten03} compared the time
variability of Z sources with that of atoll sources and tried to
associate variability components among these sources.  Based on
frequency-frequency plots, only the kHz QPOs and the horizontal branch
oscillations (HBO) found in the Z sources can be unambiguously
identified with atoll source components, the latter with $L_h$.
\citet{Straaten03} suggested that the LFN might be identified with
$L_{b2}$ and noted that (like in the case of $L_{b2}$) the
characteristic frequency of the LFN, when plotted versus $\nu_u$, does
not follow exactly the same relations between Z sources.  By comparing
the different frequency patterns in Figure~\ref{fig:b2z}, we find that
the behavior of the LFN component of GX~17+2, and that of $L_{b2}$ of
4U~1636--53 are similar, which might indicate that the physical
mechanism involved is the same. Perhaps this is related to the fact
that 4U~1636--53 is a relatively luminous atoll source
\citep[see][]{Ford00} while GX 17+2 may be a relatively low luminosity
Z-source \citep{Homan06}. Hence, 4U~1636--53 might be relatively close
in $L_x$ to GX~17+2 and differ more in $L_x$ from the other two
sources introduced above. Note that the time variability of GX~17+2 is
different from that of the other Z sources plotted in
Figure~\ref{fig:b2z}.  For instance, the characteristic frequency of
its LFN is rather low and it appears as a peak, it shows a flaring
branch oscillation (FBO) and the harmonic of the HBO is relatively
strong, whereas the other Z sources plotted show a flat LFN, no FBO
and a weak harmonic to the HBO \citep[][]{Jonker02b}.  As previously
noted by \citet{Kuulkers97}, these properties set GX~17+2 apart from
the 'Cyg-like' Z sources GX~5--1, GX~340+0 and Cyg~X-2 and, associate
it with the 'Sco-like' Z sources Sco~X-1 and GX~349+2, not plotted in
Figure \ref{fig:b2z} because no systematic study of the of the LFN and
QPO behavior of these sources in terms of $\nu_{max}$ is available as
yet.

 We further investigated the frequency similarities between
4U~1636--53 and GX~17+2 by plotting our results for the two sources.
No clear component associations were
found.
GX~17+2 is the only Z source that had
shown an anti-correlation between the frequency of one of its
components (HBO) and the kHz QPOs at high $\nu_u\gtrsim1050$~Hz
\citep{Homan02}. 
A similar effect was seen in the atoll source 4U~0614+09 between
$\nu_b$ and $\nu_u$ \citep{Straaten02}.  As can be seen in
Figure~\ref{fig:nuvsnu}, a similar decrease of $\nu_b$ with $\nu_u$ at
high frequency may occur in 4U~1636--53. However, the error bars on
$\nu_b$ are rather large in the relevant range, and the data are still
consistent with $\nu_b$ being constant at $\nu_u\gtrsim1100$~Hz, and
marginally, even with a further increase in frequency.  It is
interesting to note, that while 4U~0614+09 has a much lower $L_x$ than
4U~1636--53, both sources might show this same turnover in $\nu_b$
versus $\nu_u$.  Of course, these results need confirmation.

\subsection{The low frequency QPO}

With respect to the low-frequency QPO $L_{LF}$, 
\citet{Straaten03,Straaten05} observed that in their data there were
two groups of sources, one where the $L_{LF}$ feature was visible, and
a second one, were a QPO was detected which
they suggested to be the sub-harmonic of $L_{LF}$ and therefore,
called $L_{LF/2}$.  Following \citet{Straaten03}, the upper continuous
line in Figure~\ref{fig:bumps} indicates a power law fitted to the
$\nu_{LF}$ versus $\nu_h$ relation of the low luminosity bursters
1E~1724--3045 and GS~1826--24, and the BHC GX~339--4.  If we
reproduce the fit where we take into account the errors in both axes,
we find a best fit power-law index $\alpha = 0.97\pm0.01$ and
$\chi^2/dof = 80/19 \sim 4.2$. If we fix $\alpha=1$, the fit gives a
$\chi^2/dof=83/20 \sim 4.1$.  According to the F-test for additional
terms, there is a $<1\sigma$ improvement of the fit when $\alpha$ is
set free, so we conclude that $\nu_{LF}$ is consistent with being
linearly related to $\nu_{h}$. The lower dashed line is a
power law with the same index $\alpha=0.97$, but with a normalization half
of that of the dashed line. 

As can be seen in Figure~\ref{fig:bumps}, in 4U~1636--53 the
$L_{LF}$ component does not follow either of the two power-law
fits. If we fit the points for 4U~1636--53, we find that the power-law
index is $\alpha_2 =1.40\pm0.09$ ($\chi^2/dof = 0.14/2$),
significantly different from that of the other sources.  Given the
above, it is probably incorrect to think that the
difference in the $\nu_{LF}$ vs. $\nu_h$ relation between GX~339--4,
GS~1826--24 and 1724--3045 on one hand and 4U~1608--52, Cyg X-1 and
XTE J0929-314 on the other is associated with harmonic mode switching
\citep{Straaten03}. 
This conclusion is supported by the work of \citet{Manu05} who also
found a different correlation ($\alpha=0.58\pm0.06$, see also
Figure~\ref{fig:bumps}) in XTE~1807--294 over a much wider range of
frequencies than we obtained for 4U~1636--53,
by the high $\chi^2/dof$ for the $\nu_{LF}-\nu_h$ fit on the data of
the low luminosity bursters 1E~1724--3045, GS~1826--24, and the BHC
GX~339--4 (see previous paragraph), by the fact that if we use the
centroid frequencies instead of $\nu_{max}$, the relations worsen
\citep[see][]{Straaten03}, and by the fact that the points for
4U~1728--38 \citep{Straaten02} fall in between the two power laws,
(solid and dashed line in Figure~\ref{fig:bumps}).  Nevertheless it is
interesting that the data for 4U~0614+09, 4U~1728--34, 4U~1636--53,
4U~1820--30, 4U~1608--52 XTE~1807--294, SAX~J1808.4--3658,
XTE~J1814--338 and XTE~J0929--314 all fall on, or in between, the two
previously defined power laws, i.e., 
%
%
do not deviate from a single relation by more than a factor of 2.
We note that all the $\nu_{LF}$ values
discussed here could in principle have been affected by smearing in
the averaging process discussed in
Section~\ref{sec:dataanalysis}. However, for smearing to shift a
frequency-frequency point away from its proper value, large systematic
differences are required between the two components in the dependence
of amplitude or Q on frequency, and in the case of $L_{LF}$ and
$L_{h}$ there is no evidence for this.

From Figure~\ref{fig:bumps} it is apparent that the frequency range in
which the $L_{LF}$ component has been identified is rather large (up
to 2.5 decades).  Clearly, which frequency ranges $L_{LF}$ covers is
not related to source spin frequency,
angular momentum or luminosity of the object.  
The sources 4U~1608--52, 4U~1820--30, 4U~1636--53 and 4U~1728--34 all show
$L_{LF}$ when they are in their island state, but with $\nu_{LF}$ 
$\lesssim2.6$~Hz for 4U~1608--52, and $\gtrsim30$~Hz for the
other three sources.
The accreting millisecond pulsar XTE~J1807--294 shows $\nu_{LF}
\gtrsim 12$~Hz while the AMP XTE~J0929--314 shows $\nu_{LF} \lesssim
1$~Hz, while both have very similar spin frequencies (191~Hz,
\citealt{Markwardt03C} and 185~Hz, \citealt{Remillard02C},
respectively).  4U~1820--30 and 4U~1636--53 are at least one order of
magnitude more luminous than XTE~1807--294 and SAX~J1808.4--3658 at
their brightest \citep[see][]{Ford00,Wijnands05}, but all four sources
show $\nu_{LF} > 5$~Hz.  The only systematic feature in the LF QPO
frequencies is that while frequencies up to 50~Hz are seen in neutron
stars, black holes have not been reported to exceed 3.2~Hz, nor did
atoll sources in the EIS exceed 2.6 Hz.
So, BHCs and NS in the extreme island state are similar in this
respect; (this may be related to an overall similarity in power
spectral shape for such sources in these states that was noted before;
see, e.g., \citealt{Psaltis99,Nowak00,Belloni02,Straaten02}).

\subsection{The X-ray luminosity dependence of rms}

It has been suggested that an anti-correlation may exist between the
average X-ray luminosity of different sources and the rms amplitude of
their power spectral components \citep[see discussion in][and
references therein]{Jonker01,Straaten02,Straaten03}.  From
Figure~\ref{fig:rmsvsnu} we find differences in kHz QPO rms amplitudes
of no more than a factor 2 between sources which differ in average
luminosity by a factor up to 10, except for one point of 4U~0614+09 at
$\nu_u \sim 1140$~Hz, where the rms of the upper kHz QPOs is a factor
$\sim7$ higher than that of the other atoll sources.
\citet{Mendez01}, \citet{Jonker01} and \citet{Straaten02} have already
noted that the data are inconsistent with a model in which the
absolute amplitudes of the kHz QPOs are the same among sources, and
the decrease in rms with luminosity between sources is only caused by
an additional source of X-rays unrelated to the kHz QPOs.

From Figure~\ref{fig:rmsvsnu} it can also be seen that the largest rms
amplitude differences are found in the hHz QPOs (excluding
4U~1608--52, which is a transient source covering a large $L_x$
range). For this component we find
(1) 4U~0614+09, which has the strongest $rms_{hHz}$ ($>15$\% when
    $\nu_u\lesssim 800$~Hz);
(2) 4U~1636--53, which has the weakest $rms_{hHz}$  ($<10$\% when
    $\nu_u\lesssim 800$~Hz)
and (3) 4U~1728--34 which has $rms_{hHz}$ generally between those of
(1) and (2) [between 10 and 15\% when $\nu_u\lesssim 800$~Hz].  (At
$\nu_u\gtrsim 800$~Hz, the groups can still be differentiated as the
rms amplitude decreases with $\nu_u$).
From figure 1 in \citet{Ford00}, it can be seen that 4U~0614+09 is the
faintest X-ray source of our sample, while 4U~1636--53 is the
brightest.  4U~1728--34 show luminosities between the first two.  This
suggests an X-ray luminosity--rms anti-correlation for $L_{hHz}$ that
is not as clear in the other components (see also Figure~\ref{fig:LF}).

The fact that the rms of $L_{hHz}$ starts to decrease at the same
$\nu_u$ as that of $L_u$ and $L_b$, while $\nu_{hHz}$ does not
correlate with $\nu_u$ as all other frequencies do, suggests that the
frequency setting mechanism is different for $L_{hHz}$ compared with
the other components, while the amplitude setting mechanism is common.
As pointed out in Section~\ref{sec:results}, the drop in rms in $L_u$,
$L_{hHz}$ and $L_b$ starts at $\nu_u$ between $700$ and $800$~Hz. For
the case of 4U~1636--53 shown here, this corresponds to interval
G. The power spectrum of this interval may represent the transition
between the island and the banana state, when the geometric
configuration of the system is thought to change
\citep[e.g.][]{Jonker00b,Gierlinski02}. For example, the appearance of
a puffed-up disk could smear out the variability coming from the inner
region where the oscillations are produced.
\subsection{The nature of the hectohertz QPOs}

While our results indicate that the characteristic frequency
of the hHz QPO may oscillate as a function of $\nu_u$, $\nu_{hHz}$
remains constrained to a limited range of frequencies (100--250~Hz) for
4U~1636--53 and for the other sources used in Figure~\ref{fig:nuvsnu}.
A similar result has been reported for $\nu_{hHz}$ in several other
atoll sources such as in MXB~1730--335 \citep{Migliari05}, 4U~1820--30
\citep{Altamirano05} and in the atoll source and millisecond accreting
pulsar SAX~J1808.4--3658 \citep{Wijnands98,Straaten05}.  Interestingly,
the presence of $L_{hHz}$ has not been confirmed for Z-sources
\citep{Vanderklis06}, possibly due to the intrinsic differences
between atoll and Z-sources such as  luminosity.

\citet{Straaten02} have suggested a link between the
$\lesssim100$~Hz QPOs reported by \citet{Nowak00} in the black holes
Cyg~X-1 and GX~339--34 and $L_{hHz}$. \citet{Straaten02} also suggested
that $L_{hHz}$ could be related to the $\sim67$~Hz QPO in the black
hole GRS 1915+105 \citep{Morgan97} and the $\sim300$~Hz QPO in the BHC
GRO J1655--40 \citep{Remillard99b} which also have stable frequencies.
\citet{Fragile01} made a tentative identification of the $\sim9$~Hz
QPO in the BHC GRO~J1655--40 \citep{Remillard99b} with the orbital
frequency at the Bardeen-Petterson (B--P) transition radius
\citep{Bardeen75} and suggested the same identification for $L_{hHz}$
in neutron star systems.
In this scenario, the orbital
frequency at the radius where a warped disk is forced to the
equatorial plane by the Bardeen--Petterson 
 effect can produce a quasi-periodic signal \citep[see][for
an schematic illustration of the scenario]{Fragile01}.

Attempts have been made to theoretically 
estimate the B--P transition radius from accretion disks models in terms of 
the angular momentum and the mass of the compact object
\citep[e.g.][]{Bardeen75,Ivanov97,Hatchett81,Nelson00}.  
\citet{Fragile01} propose a parameterization involving
a scaling parameter A, which according to them  lies in the range
$10\lesssim A \lesssim300$.  These authors write the B--P radius as 
$R_{BP}=A \cdot a_{\star}^{2/3} \cdot R_{GR}$, where
$a_{\star}=Jc/GM^2$ is the dimensionless specific angular momentum
(J and M are the angular momentum and the mass the compact
object, respectively) and $R_{GR}$ is $GM/c^2$. The 
Keplerian orbital frequency associated with the B--P transition radius
can be written as $\nu_{kep,BP}= c^3 \cdot (2 \pi G)^{-1} \cdot (M
a_{\star} A^{3/2})^{-1}$.
If we assume that the atoll sources plotted in Figure~\ref{fig:nuvsnu}
all have masses between $1.4$ and $2M_{\odot}$, that
$0.3<a_{\star}<0.7$ \citep[see e.g.][ and references
within]{Salgado94,Cook94} and that the central frequency of the hHz
QPOs is between $\sim100$ and $\sim250$~Hz, we can constrain the
scaling factor A for these source to be between $\sim20$ and
$\sim84$. If A only depends on the accretion disk (i.e. does not
depend on the central object), this 
can be used to constrain the frequency range in which we expect to observe
 $\nu_{kep,BP}$ in black holes.
For example, the black hole BHC GRO~J1655--40, whose mass is estimated
from optical and infrared investigations as $M=6.3\pm0.5M_{\odot}$
\citep{Greene01} and whose specific angular momentum $a_{\star}$ can
be estimated to be between 0.5 and 0.95 \citep{Cui98,Fragile01}, would
have $\nu_{kep,BP}$ between $\sim6.5$ and $\sim127$~Hz, which would
exclude the 300~Hz QPO observed in GRO~J1655--40 but would be
consistent with the the 9~Hz QPO as proposed by \citet{Fragile01}.
If one assumes the 450~Hz QPO in GRO~J1655--40
is associated with orbital motion at the last stable orbit, then
$a_{\star}$ could be as low as $\sim0.15$ \citep{Strohmayer01a}. In
this case, $\nu_{kep,BP}$ can be as high as $\sim425$~Hz for a black
hole mass of $5.7 M_{\odot}$.

\section{Summary}

\begin{itemize}

\item Our observations of 4U~1636--53, including the first RXTE
island state data of the source, show timing behavior remarkably
similar to that seen in other atoll NS-LMXBs. 
We observe all components previously identified in those sources, and
find their frequencies to follow similar relations to those previously
observed. This is interesting as the sources compared in this work
were observed at intrinsic luminosities different by more than an
order of magnitude.
\item The previously proposed interpretation of the QPO frequencies
 $\nu_{LF}$ and $\nu_{LF2}$ in different sources in terms of harmonic
 mode switching is not supported by our data on 4U~1636--53, nor by
 data previously reported for other sources. However, these
 frequencies still do not deviate from a single relation by more than
 a factor of 2 for all sources.
%
%
%
%
\item The low frequency QPO $L_{LF}$ is seen in black holes and in
accreting millisecond pulsars as well as in non-pulsing neutron stars
at frequencies between $\sim0.1$ and $\sim50$~Hz. The frequency range
that $L_{LF}$ covers in a given source is not related to spin
frequency, angular momentum or luminosity of the object.
\item The rms and frequency behavior of the hectohertz QPO
suggests that the mechanism that sets its frequency differs from that
for the other components, while the amplitude setting mechanism is
common.  
%
%
%
\end{itemize}

\textbf{Acknowledgments:} DA wants to thank S. van Straaten for all
his help in the analysis of these data. DA also wants to thank
R. Wijnands and C. Fragile for very helpful comments and discussions
and J. Homan for comments on an earlier version of this manuscript.
This work was supported by the ``Nederlandse Onderzoekschool Voor
Astronomie'' (NOVA), i.e., the ``Netherlands Research School for
Astronomy'', and it has made use of data obtained through the High
Energy Astrophysics Science Archive Research Center Online Service,
provided by the NASA/Goddard Space Flight Center.

\textbf{Appendix}

In this Appendix we further discuss other possible approaches to
analyze the characteristics of complex power spectra such as generally
found in neutron star low-mass X-ray binaries.

In the ideal case, we would have data with enough statistics to be
able to follow the evolution of the parameters of all observable
components in the power spectra on sufficiently short time scales to
be sensitive to the smallest meaningful variations. Unfortunately,
this is not the case for the present data and meaningful variations are
averaged out in our data. These variations can sometimes be recovered by
the use of alternative methods. 
For example, with the ``shift and add'' method introduced by
\citet{Mendez98a}, it has been possible to better constrain some of
the characteristics of the kHz QPOs in several sources than without
this method \citep[e.g.][]{Mendez98a,Barret05a}. A disadvantage of the
method is that it distorts the power spectrum at the lowest and
highest frequencies covered.

We investigated if this method could also be used for our
purpose. However, in our experiments with this we encountered several
complications.
From the observational point of view, in order to use this method we
require a sharp power spectral feature that can be accurately traced
in time. There are two possibilities for such features: the lower kHz
QPO $L_{\ell}$ and the low-frequency Lorentzian $L_{LF}$. While the
lower kHz QPO is usually superimposed on well-modeled Poisson noise,
tracing $L_{LF}$ is complicated by the fact that it is 
superimposed on strong variable broad band noise \citep[see][ and
references within]{Vanderklis06}.
More importantly, while $L_{\ell}$ can be traced on sufficiently short
time scales ($\lesssim64$sec) for the intrinsic changes in the
characteristics of the power spectrum to be minimal, typically an
entire observation is required for detecting $L_{LF}$ at sufficient
signal to noise.
In practice this means that we can only use the shift and add
method with $L_{\ell}$, which constrains us to only that relatively
limited part of the data where $L_{\ell}$ is actually detected (see
Figure~\ref{fig:nuvsnu}). 
We note that $L_{LF}$ and
$L_{\ell}$ are not simultaneously detected in our data set.

We analyzed all the datasets described in
Section~\ref{sec:dataanalysis} and found that $\sim15$\%
($\sim0.17$~Msec) of our data have traceable lower kHz QPOs. Most of
that time the lower kHz QPOs are detected at frequencies between 700
and 850 Hz (the full range was 600--900~Hz).

In order to use the shift and add method we must adopt a relation
between the frequency of the component we wish to shift on (here the
lower kHz QPO) and the frequency of the component we wish to detect
(here the low frequency QPOs/noise).  For example, in their original
work \citet{Mendez98a} supposed that the difference between the lower
and the upper kHz QPO frequency remained constant when both peaks move.
The study of the characteristics of the low frequency QPOs/noise using
the shift and add method is complicated by the fact that we have
imprecise information about their relation with the lower kHz QPOs: a
constant frequency difference certainly does not apply even to narrow
ranges in shift frequency. The aim of this paper as well as the aim of
the papers cited below is to present observational results that help
constrain those relations.

As \citet{Straaten02,Straaten03} showed, the frequencies of all
components except those of the hHz QPOs are correlated in a similar
way between sources (see Figure~\ref{fig:nuvsnu}). However,
\citet{Straaten05} and \citet{Manu05} also showed that those
correlations are shifted in pulsating sources and \citet{Altamirano05}
found that even non-pulsating systems might show frequency shifts.
Additionally, the results of \citet{Straaten02,Straaten03,Straaten05},
\citet{Manu05} and \citet{Altamirano05} show that although the
frequency relations between the different components are well fitted
with a power law, the index and normalization of the power law are
different for each relation and may also depend on frequency range.

In order to quantify the problem described above, we studied
observation 60032-01-05-00 using power spectra of 64-sec data segments
at 2~Hz frequency resolution. This is a very good observation for our
purpose since:
(i) it has $\sim27$~ksec of uninterrupted data;
(ii) the lower kHz QPO is strong enough to be significantly detected
within 64 seconds for the entire 27~ksec;
(iii) the lower kHz QPO frequency drifts between $\sim700$ and $\sim860$~Hz and
(iv) the power spectrum can be fitted with 4 Lorentzians: 2 for the
kHz QPOs, one for $L_b$ (at $40\pm4$~Hz) and one for $L_{b2}$ (at
$20\pm12$~Hz; $2.9\sigma$).

We first analyzed the power spectrum obtained by aligning the
$L_{\ell}$ components. We found that $L_b$ and $L_{b2}$ had blended
into a broad component at $\sim100$~Hz. This result was expected, as a
drift of 160~Hz in $\nu_{\ell}$ does not imply a drift of the same
magnitude in the frequencies of the low $\nu$ components.
We then tried to align the power spectra by predicting the position of
the low-frequency components from $\nu_{\ell}$ using a different power
law relation for each component as reported by \citet[][]{Straaten05}
for $L_b$ and $\nu_{b2} =  e^{-24\pm8} \cdot \nu_u^{+3.8\pm1.3}$ for
$L_{b2}$ (the relation for $L_{b2}$ is based on our data for
4U~1636--53; given the large errors in our data, the $\chi^2/dof$ for
the power law fit was 0.26) .
The results  depended on which power law we used: no
significant changes in the resulting power spectrum were found when
trying to align $L_{b2}$, while the power of both
components was smeared out producing a blend when we tried to align
$L_b$. This was clearly the effect of the difference
in power law indices and normalizations between components. 
As the frequency relations we used are between $\nu_u$ and
the frequencies of the other components, we had to assume a relation
between $\nu_{\ell}$ and $\nu_u$ in order to predict the frequency
variations. We variously assumed $\Delta\nu=\nu_u-\nu_{\ell}=280,\ 300
\ \& \ 350$~Hz and obtained similar results in each case. We also
predicted $\Delta\nu$ by fitting a line to the $\Delta\nu$
vs. $\nu_{\ell}$ data reported by \citet{Jonker02} in the range
$720\lesssim \nu_{\ell} \lesssim 900$~Hz. Again, the results of the
power-spectral fits were the same within errors as those of the
previous experiments.

We repeated the last exercise (using the power law relations) also for
all 0.17~Msec of $\nu_{\ell}$ useful data and for $L_{b2}$, $L_b$ and
$L_h$ (we use the power law relation as reported by \citealt{Straaten05} for
$L_h$). We again found that our results were dependent on the power
law used and not significantly better defined than the average power
spectra obtained when all 0.17~Msec of data were averaged together
without shift.

In another experiment we calculated 4 average power spectra including
all 0.17~Msec of useful data by selecting only those 64-second
segments which had $\nu_{\ell}$ between 650--700, 700--750, 750--800,
and 800--850~Hz respectively, and averaging these selected power
spectra without shifting. In all cases we detect both kHz QPOs, a
power law VLFN and $L_b$. As expected from
the results shown in Figure~\ref{fig:nuvsnu}, $\nu_b$ is
correlated with the frequency of both kHz QPOs. The measured
frequencies are all consistent within errors with those reported on
Figure~\ref{fig:nuvsnu} and Table~\ref{table:data}. The lack of
statistics in each average power spectrum did not allow us to well
constrain the power spectral parameters of other components.

Finally, we fitted a line to the relation between $\nu_{LF}$ and
$\nu_{h}$ defined by the 4-points visible in
Figure~\ref{fig:bumps}. We used the $\nu_{LF}$ we find in all four
power spectra  (A1, B, D and E) to predict $\nu_h$
and shift and add these four power spectra together.  Again, we find
that the blend of components (this time between $L_b$ and $L_h$) and
the distortion of the power spectra at low frequencies prevented us to
better estimate the $L_h$ parameters.
So, neither the shift and add method nor selecting data on $\nu_{\ell}$
in 64-sec segments (i.e. much shorter than an observation) in our data
provides an advantage in measuring the broad low frequency components
better. Therefore, in this paper we decided to the straightforward
method described in Section~\ref{sec:dataanalysis}.

\clearpage

\begin{longtable}{cccc} 
\caption{Characteristic frequencies $\nu_{max}$, $Q$ values
($\nu_{max} \equiv \nu_{central}/FWHM$), fractional rms (in the full
PCA energy band) and component identification (ID) of the Lorentzians
fitted for 4U~1636--53.  The quoted errors use $\Delta\chi^2 =
1.0$. Where only one error is quoted, it is the straight average
between the positive and the negative error. Note that the results for
the quality factor of both $L_{\ell}$ and $L_{LF}$ components are
affected by our averaging method (See Sections~\ref{sec:intro},
\ref{sec:results} and the Appendix)} \label{table:data}

\endfirsthead

\hline \multicolumn{4}{|r|}{{Continued on next Column}} \\ \hline
\endfoot

\hline \hline
\endlastfoot

\multicolumn{4}{c}{Interval A1}\\ 
\vspace{0.1cm} \\
$\nu_{max}$ (Hz)& $Q$          & RMS (\%)      & ID\\ 
\hline \\
$434.8\pm 26.1$ & $1.6\pm 0.5$ & $12.9\pm 1.4$ & $L_u$ \\ 
$175.9\pm 8.9$ & $2.7\pm 1.3$ & $8.06\pm 1.23$ & $L_{hHz}$ \\ 
$15.0\pm 0.6$ & $0.7\pm 0.1$ & $13.3\pm 0.7$ & $L_h$ \\ 
$5.1\pm 0.1$ & $7.3^{+9.0}_{-2.2}$ & $3.2\pm 0.6$ & $L_{LF}$ \\ 
$3.2\pm 0.4$ & $0.33\pm 0.07$ & $11.1\pm 0.7$ & $L_b$ \\ 
\hline \\ 
\vspace{0.2cm} \\
\multicolumn{4}{c}{Interval A2}\\ 
\vspace{0.1cm} \\
$\nu_{max}$ (Hz) & $Q$ & RMS (\%)&  ID \\
\hline \\
$436.4\pm 28.2$ & $0.92\pm 0.24$ & $16.5\pm 1.05$ & $L_u$ \\ 
$12.3\pm 0.8$ & $0.27\pm 0.11$ & $16.1\pm 1.1$ & $L_h$ \\ 
$2.34\pm 0.52$ & $0.14\pm 0.09$ & $9.9\pm 1.4$ & $L_b$ \\ 
\hline \\ 
\vspace{0.2cm} \\
\multicolumn{4}{c}{Interval B}\\
\vspace{0.1cm} \\
$\nu_{max}$ (Hz) & $Q$ & RMS (\%)&   ID\\
\hline \\
$464.9\pm 9.4$ & $1.7\pm 0.2$ & $13.8\pm 0.9$ & $L_u$ \\ 
$154.7^{+68.8}_{-36.6}$ & $0.53\pm 0.42$ & $7.4\pm 1.8$ & $L_{hHz}$ \\ 
$18.2\pm 0.3$ & $0.75\pm 0.06$ & $12.4\pm 0.4$ & $L_h$ \\ 
$6.83\pm 0.11$ & $3.07\pm 0.66$ & $4.1\pm 0.5$ & $L_{LF}$ \\ 
$4.03\pm 0.31$ & $0.19\pm 0.03$ & $11.9\pm 0.4$ & $L_b$ \\ 
\hline \\ 
\vspace{0.2cm} \\
\multicolumn{4}{c}{Interval C}\\ 
\vspace{0.1cm} \\
$\nu_{max}$ (Hz) & $Q$ & RMS (\%)&  ID \\
\hline \\
$529.3\pm 15.4$ & $1.1\pm 0.1$ & $17.1\pm 0.6$ & $L_u$ \\ 
$23.1\pm 1.6$ & $0.47\pm 0.15$ & $11.9\pm 1.5$ &  $L_h$ \\ 
$6.38\pm 1.01$ & $0.09\pm 0.05$ & $13.1\pm 1.2$ & $L_b$ \\ 
\hline \\ 
\vspace{0.2cm} \\
\multicolumn{4}{c}{Interval D}\\ 
\vspace{0.1cm} \\
$\nu_{max}$ (Hz) & $Q$ & RMS (\%)&  ID \\
\hline \\
$524.7\pm 8.0$ & $2.2\pm 0.4$ & $13.9\pm1.0$ & $L_u$ \\ 
$201.7^{+78.0}_{-32.5}$ & $0.58\pm 0.35$ & $8.6\pm1.9$ & $L_{hHz}$ \\ 
$23.9\pm 0.5$ & $0.99\pm 0.13$ & $10.5\pm0.6$ & $L_h$ \\ 
$9.8\pm 0.3$ & $2.17\pm 0.63$ & $5.2\pm0.9$ & $L_{LF}$ \\ 
$5.6\pm 0.7$ & $0.15\pm 0.04$ & $11.4\pm0.7$ &  $L_b$\\
\hline \\ 
\vspace{0.2cm} \\
\multicolumn{4}{c}{Interval E}\\ 
\vspace{0.1cm} \\
$\nu_{max}$ (Hz) & $Q$ & RMS (\%)&  ID \\
\hline \\
$593.1\pm 4.3$ & $3.1\pm 0.2$ & $12.7\pm0.4$ & $L_u$ \\ 
$270.3\pm 31.7$ & $0.63\pm 0.20$ & $8.7\pm0.8$ & $L_{hHz}$ \\ 
$31.2\pm 1.3$ & $0.89\pm 0.15$ & $8.7\pm0.7$ &  $L_h$\\ 
$15.04\pm 0.41$ & $1.6\pm 0.5$ & $5.7\pm1.2$ &  $L_{LF}$\\ 
$8.5\pm 0.9$ & $0.22\pm 0.03$ & $11.2\pm0.6$ & $L_b$\\ 
\hline \\ 
\vspace{0.2cm} \\
\multicolumn{4}{c}{Interval F}\\ 
\vspace{0.1cm} \\
$\nu_{max}$ (Hz) & $Q$ & RMS (\%)&  ID \\
\hline \\
$637.2\pm 7.9$ & $3.06\pm 0.32$ & $15.9\pm0.6$ & $L_u$ \\ 
$208.7\pm 30.1$ & $1.2\pm 0.6$ & $7.2\pm1.2$ & $L_{hHz}$ \\ 
$18.7\pm 0.9$ & $0.02\pm 0.04$ & $16.9\pm0.2$ &  $L_b$ \\ 
\hline \\ 
\vspace{0.2cm} \\
\multicolumn{4}{c}{Interval G}\\ 
\vspace{0.1cm} \\
$\nu_{max}$ (Hz) & $Q$ & RMS (\%)&  ID \\
\hline \\
$780.01\pm 1.78$ & $4.5\pm 0.1$ & $13.6\pm 0.1$ &  $L_u$ \\ 
$152.7\pm 10.4$ & $0.32\pm 0.11$ & $8.7\pm 0.49$ &  $L_{hHz}$ \\ 
$23.4\pm 0.5$ & $0.34\pm 0.03$ & $11.6\pm 0.3$ &  $L_b$ \\ 
$3.8\pm 0.9$ & $0.20\pm 0.11$ & $2.8\pm 0.5$ &  $L_{b2}$ \\ 
\hline \\ 
\vspace{0.2cm} \\
\multicolumn{4}{c}{Interval H}\\ 
\vspace{0.1cm} \\
$\nu_{max}$ (Hz) & $Q$ & RMS (\%)&  ID \\
\hline \\
$862.8\pm 1.8$ & $6.2\pm 0.2$ & $11.4\pm0.1$ & $L_u$ \\ 
$568.3\pm 4.6$ & $7.06\pm 1.22$ & $4.6\pm0.2$ & $L_{\ell}$ \\ 
$120.6\pm 3.6$ & $1.03\pm 0.14$ & $6.4\pm0.3$ & $L_{hHz}$ \\ 
$26.6\pm 0.5$ & $0.57\pm 0.04$ & $9.2\pm0.2$ & $L_b$ \\ 
$4.3\pm 0.7$ & $0.35\pm 0.11$ & $2.7\pm0.3$ & $L_{b2}$ \\
\hline \\ 
\vspace{0.2cm} \\
\multicolumn{4}{c}{Interval I}\\ 
\vspace{0.1cm} \\
$\nu_{max}$ (Hz) & $Q$ & RMS (\%)&  ID \\
\hline \\
$897.9\pm 2.1$ & $8.2\pm 0.5$ & $9.9\pm 0.2$ &  $L_u$\\ 
$585.5\pm 5.3$ & $7.6\pm 1.1$ & $5.3\pm 0.3$ &  $L_{\ell}$\\ 
$228.81\pm 9.05$ & $5.6^{+5.6}_{-1.9}$ & $2.6\pm 0.5$ & $L_{hHz}^{harmonic}$ \\ 
$112.68\pm 4.34$ & $1.51\pm 0.33$ & $5.29\pm 0.41$ & $L_{hHz}$ \\ 
$28.6\pm 0.6$ & $0.80\pm 0.09$ & $7.72\pm 0.28$ & $L_b$ \\ 
$6.06\pm 1.41$ & $0.32\pm 0.14$ & $2.93\pm 0.41$ &  $L_{b2}$\\ 
\hline \\ 
\vspace{0.2cm} \\
\multicolumn{4}{c}{Interval J}\\ 
\vspace{0.1cm} \\
$\nu_{max}$ (Hz) & $Q$ & RMS (\%)&  ID \\
\hline \\
$971.1\pm 4.7$ & $9.9\pm 1.1$ & $6.7\pm 0.2$ & $L_u$ \\ 
$666.2\pm 2.1$ & $8.7\pm 0.5$ & $7.7\pm 0.2$ & $L_{\ell}$ \\ 
$127.9\pm 9.6$ & $1.1\pm 0.3$ & $4.3\pm 0.4$ & $L_{hHz}$ \\ 
$34.6\pm 0.8$ & $1.1\pm 0.1$ & $5.5\pm 0.2$ & $L_b$ \\ 
$8.07\pm 1.56$ & $0.34\pm 0.14$ & $2.99\pm 0.31$ &  $L_{b2}$\\ 
\hline \\ 
\vspace{0.2cm} \\
\multicolumn{4}{c}{Interval K}\\ 
\vspace{0.1cm} \\
$\nu_{max}$ (Hz) & $Q$ & RMS (\%)&  ID \\
\hline \\
$998.18\pm 7.94$ & $5.76\pm 0.74$ & $6.14\pm 0.23$ & $L_u$ \\ 
$728.04\pm 2.63$ & $5.3\pm 0.3$ & $8.3\pm 0.2$ &  $L_{\ell}$\\ 
$148.2\pm 8.1$ & $1.6\pm 0.5$ & $3.4\pm 0.3$ & $L_{hHz}$ \\ 
$38.8\pm 0.9$ & $1.2\pm 0.2$ & $4.69\pm 0.37$ & $L_b$ \\ 
$9.5^{+12}_{-3.2}$ & $0.07\pm 0.35$ & $2.48\pm 0.79$ & $L_{b2}$ \\ 
\hline \\ 
\vspace{0.2cm} \\
\multicolumn{4}{c}{Interval L}\\ 
\vspace{0.1cm} \\
$\nu_{max}$ (Hz) & $Q$ & RMS (\%)&  ID \\
\hline \\
$1138.37\pm 9.02$ & $14.08^{+9.80}_{-3.66}$ & $2.7\pm 0.3$ &  $L_u$ \\ 
$837.5\pm 1.0$ & $11.4\pm 0.4$ & $7.9\pm 0.09$ &  $L_{\ell}$ \\ 
$156.1\pm 33.6$ & $1.3\pm 0.5$ & $2.03\pm 0.38$ &  $L_{hHz}$ \\ 
$46.4\pm 0.9$ & $2.4\pm 0.4$ & $2.8\pm 0.2$ &  $L_{b}$\\ 
$14.9\pm 5.8$ & $0.00\pm 0.00$ & $1.9\pm 0.3$ &  $L_{b2}$\\ 
\hline \\ 
\vspace{0.2cm} \\
\multicolumn{4}{c}{Interval M}\\ 
\vspace{0.1cm} \\
$\nu_{max}$ (Hz) & $Q$ & RMS (\%)&  ID \\
\hline \\
$1220.5\pm 11.2$ & $16.6\pm 6.5$ & $3.71\pm 0.34$ & $L_u$ \\ 
$44.2\pm 7.2$ & $0.56\pm 0.28$ & $3.1\pm 0.3$ & $L_b$ \\ 
\hline \\ 
\vspace{0.2cm} \\
\multicolumn{4}{c}{Interval N}\\ 
\vspace{0.1cm} \\
$\nu_{max}$ (Hz) & $Q$ & RMS (\%)&  ID \\
\hline \\
$1259.1\pm 9.9$ & $14.4^{+34.7}_{-5.4}$ & $1.8\pm0.2$ & $L_u$ \\ 
$36.7\pm 4.9$ & $0.05\pm 0.13$ & $2.37\pm0.11$ & $L_b$ \\ 
\end{longtable}

\clearpage


\end{document}